\documentclass[superscriptaddress,preprint,showpacs,preprintnumbers,amsmath,amssymb]{revtex4}

\usepackage{graphicx}
\usepackage{dcolumn}
\usepackage{bm}

\begin{document}


\title{Polarization transfer measurements for 
$^{12}{\rm C}(\vec{p},\vec{n})^{12}{\rm N (g.s.},1^+)$ at 296~MeV 
and nuclear correlation effects}

\author{M. Dozono}
\email{dozono@phys.kyushu-u.ac.jp}
\affiliation{Department of Physics, Kyushu University, Higashi, 
Fukuoka 812-8581, Japan}
\author{T. Wakasa}
\affiliation{Department of Physics, Kyushu University, Higashi, 
Fukuoka 812-8581, Japan}
\author{E. Ihara}
\affiliation{Department of Physics, Kyushu University, Higashi, 
Fukuoka 812-8581, Japan}
\author{S. Asaji}
\affiliation{Department of Physics, Kyushu University, Higashi, 
Fukuoka 812-8581, Japan}
\author{K. Fujita}
\affiliation{Department of Physics, Kyushu University, Higashi, 
Fukuoka 812-8581, Japan}
\author{K. Hatanaka}
\affiliation{Research Center for Nuclear Physics, Osaka University,
Ibaraki, Osaka 567-0047, Japan}
\author{\\M. Ichimura}
\affiliation{Faculty of Computer and Information Sciences, 
Hosei University, Koganei, Tokyo 184-8584, Japan}
\author{T. Ishida}
\affiliation{Cyclotron and Radioisotope Center, Tohoku University, 
Aoba, Miyagi 980-8578, Japan}
\author{T. Kaneda}
\affiliation{Research Center for Nuclear Physics, Osaka University,
Ibaraki, Osaka 567-0047, Japan}
\author{H. Matsubara}
\affiliation{Research Center for Nuclear Physics, Osaka University,
Ibaraki, Osaka 567-0047, Japan}
\author{Y. Nagasue}
\affiliation{Department of Physics, Kyushu University, Higashi, 
Fukuoka 812-8581, Japan}
\author{T. Noro}
\affiliation{Department of Physics, Kyushu University, Higashi, 
Fukuoka 812-8581, Japan}
\author{\\Y. Sakemi}
\affiliation{Cyclotron and Radioisotope Center, Tohoku University, 
Aoba, Miyagi 980-8578, Japan}
\author{Y. Shimizu}
\affiliation{Center for Nuclear Study, The University of Tokyo, 
Bunkyo, Tokyo 113-0033, Japan}
\author{H. Takeda}
\affiliation{Department of Physics, Kyushu University, Higashi, 
Fukuoka 812-8581, Japan}
\author{Y. Tameshige}
\affiliation{Research Center for Nuclear Physics, Osaka University,
Ibaraki, Osaka 567-0047, Japan}
\author{A. Tamii}
\affiliation{Research Center for Nuclear Physics, Osaka University,
Ibaraki, Osaka 567-0047, Japan}
\author{Y. Yamada}
\affiliation{Department of Physics, Kyushu University, Higashi, 
Fukuoka 812-8581, Japan}

\date{\today}

\begin{abstract}
Differential cross sections and complete sets of 
polarization observables are presented for the Gamow-Teller 
$^{12}{\rm C}(\vec{p},\vec{n})^{12}{\rm N}({\rm g.s.},1^+)$ 
reaction at a bombarding energy of 296~MeV with 
momentum transfers $q$ of $0.1$ to $2.2~{\rm fm}^{-1}$. 
The polarization transfer observables are 
used to deduce the spin-longitudinal cross section, $ID_q$, 
and spin-transverse cross sections, $ID_p$ and $ID_n$. 
The data are compared with calculations based on the distorted wave impulse approximation 
(DWIA) using shell-model wave functions. 
Significant differences between the experimental and theoretical 
results are observed for all three spin-dependent $ID_i$ 
at momentum transfers of $q \gtrsim 0.5~{\rm fm}^{-1}$, 
suggesting the existence of 
nuclear correlations beyond the shell model. 
We also performed DWIA calculations employing 
random phase approximation (RPA) response functions 
and found that the observed discrepancy is partly 
resolved by the pionic and rho-mesonic correlation effects. 

\end{abstract}

\maketitle

\section{\label{sec:introduction}INTRODUCTION}
Nuclear spin--isospin correlations produce various 
interesting phenomena in nuclei 
depending on the momentum transfer $q$~\cite{ichimura2006}. 
At small momentum transfers, $q \simeq 0~{\rm fm}^{-1}$, 
the spin--isospin particle-hole interaction is strongly repulsive, 
which leads to such collective excitations in nuclei as 
the Gamow-Teller (GT) giant resonance. 
The quenching of the total strength of the GT transitions from its sum 
rule value has prompted theoretical studies of 
possible mechanisms, ranging from 
conventional configuration mixing 
to admixture of the $\Delta$-hole ($\Delta$-h) states. 
Recent experimental studies of 
$^{90}{\rm Zr}(p,n)$ and $(n,p)$ reactions~\cite{wakasa1997,yako2005} 
have revealed that the coupling to 2-particle--2-hole (2p-2h) 
excitations is the major source of quenching of the GT strengths, 
while $\Delta$-h coupling plays a minor role. 

At fairly large momentum transfers, $q \gtrsim 1~{\rm fm}^{-1}$, 
the spin-transverse interaction induced by one rho-meson 
exchange remains repulsive, while the spin-longitudinal interaction 
driven by one pion exchange becomes attractive. 
This attraction of the spin-longitudinal interaction 
produces pion condensation and its precursor phenomena. 
Pion condensation~\cite{migdal} 
is expected to occur in neutron stars (NSs) 
such as 3C58~\cite{slane2002} 
and accelerate their cooling~\cite{tsuruta2002}. 
It is predicted that pion condensation does not occur in normal nuclei. 
However, precursor phenomena may be observed in normal nuclei 
if they are near the critical point of the phase 
transition. 
As possible evidence of a precursor, 
enhancements of the M1 cross section in proton inelastic 
scattering~\cite{toki1979,delorme1980} 
and of the ratio $R_L/R_T$, 
the spin-longitudinal (pionic) response function $R_L$ 
to the spin-transverse (rho-mesonic) response function $R_T$, 
in the quasielastic scattering 
(QES) region~\cite{alberico1980,alberico1982} 
were proposed at a momentum transfer of about $q \simeq 1.7~{\rm fm}^{-1}$. 
Motivated by these predictions, 
many experiments involving the M1 transition and QES have been carried out. 
These include the measurement of the
$^{12}{\rm C}(p,p')^{12}{\rm C}^*(1^+,T=1)$ scattering 
at incident energies of about 120 to 
800~MeV~\cite{comfort1980-1,comfort1980-2,haji-saeid1980,comfort1981-1}, 
and the experimental extraction of $R_L/R_T$ using 
complete sets of polarization transfer observables 
in $(\vec{p},\vec{p}\  ')$ 
scattering~\cite{carey1984,rees1986,fergerson1988,hausser1988,chan1990} 
and $(\vec{p},\vec{n})$ 
reactions~\cite{mcclelland1992,chen1993,taddeucci1994,wakasa1999,hautala2002} 
on various targets at intermediate energies. 
However, these experimental data did not show the 
theoretically expected enhancements. 
Recent analysis of QES data shows pionic enhancement in the 
spin-longitudinal cross section representing $R_L$, 
which suggests that the lack of enhancement in the value of $R_L/R_T$ is due to 
the rho-mesonic component~\cite{wakasa2004-2}. 
The measurement of the pure pionic excitation of 
$^{16}{\rm O}(p,p')^{16}{\rm O}^*(0^-,T=1)$ 
scattering at $T_p=295~{\rm MeV}$ 
also supports such an enhancement~\cite{wakasa2006}. 

The analyses of QES and pure pionic excitation data suggest that 
one needs to reconsider the interpretation of the M1 data. 
The pionic effect in the M1 cross section might be masked by 
the contribution from the rho-mesonic component 
because the M1 state is a mixture of spin-longitudinal 
and spin-transverse states. 
In addition, proton inelastic scattering, including 
the M1 transition, mix isoscalar and isovector contributions. 
With respect to this issue, the GT 
$^{12}{\rm C}(p,n)^{12}{\rm N}({\rm g.s.},1^+)$ 
reaction, which is the isobaric analog of the M1 
$^{12}{\rm C}(p,p')^{12}{\rm C}^*(1^+,T=1)$ 
scattering, has an advantage because the $(p,n)$ reaction 
exclusively depends on the isovector contribution. 
Thus, it is interesting to study nuclear correlation effects 
in this reaction by separating the cross section into 
pionic and rho-mesonic components, using
a complete set of polarization observables. 
However, no complete polarization transfer 
measurements for the GT reaction have been reported until now. 

In this paper, we present differential cross sections and 
complete sets of polarization observables for the 
$^{12}{\rm C}(\vec{p},\vec{n})^{12}{\rm N}({\rm g.s.},1^+)$ 
reaction measured at $T_p=296~{\rm MeV}$ over a range of 
$q=0.1$ to $2.2~{\rm fm}^{-1}$. 
This incident energy is ideal for studying GT 
transitions because the spin excitations 
are dominant in the $(p,n)$ reaction near 300~MeV~\cite{franey1985}. 
In addition, distortion effects are minimal around 
300~MeV~\cite{ichimura2006}, 
thereby enabling one to extract reliable nuclear structure 
information such as nuclear correlation effects.  

The differential cross section and a complete set of polarization 
transfer observables are used to separate the cross section into 
nonspin ($ID_0$), spin-longitudinal ($ID_q$), and two 
spin-transverse ($ID_p$ and $ID_n$) polarized cross sections. 
The spin-dependent polarized cross sections, $ID_q$, $ID_p$, and $ID_n$, 
are compared with distorted wave impulse approximation (DWIA) 
calculations using random phase approximation (RPA) 
response functions, 
in order to assess the nuclear correlation effects quantitatively. 
The theoretical calculations give good descriptions of 
all of the spin-dependent polarized cross sections. 
These results demonstrate the existence of pionic and rho-mesonic 
correlations inside nuclei. 

\section{\label{sec:experimental_methods}EXPERIMENTAL METHODS}
\begin{figure}[t]
\includegraphics[width=\hsize,clip]{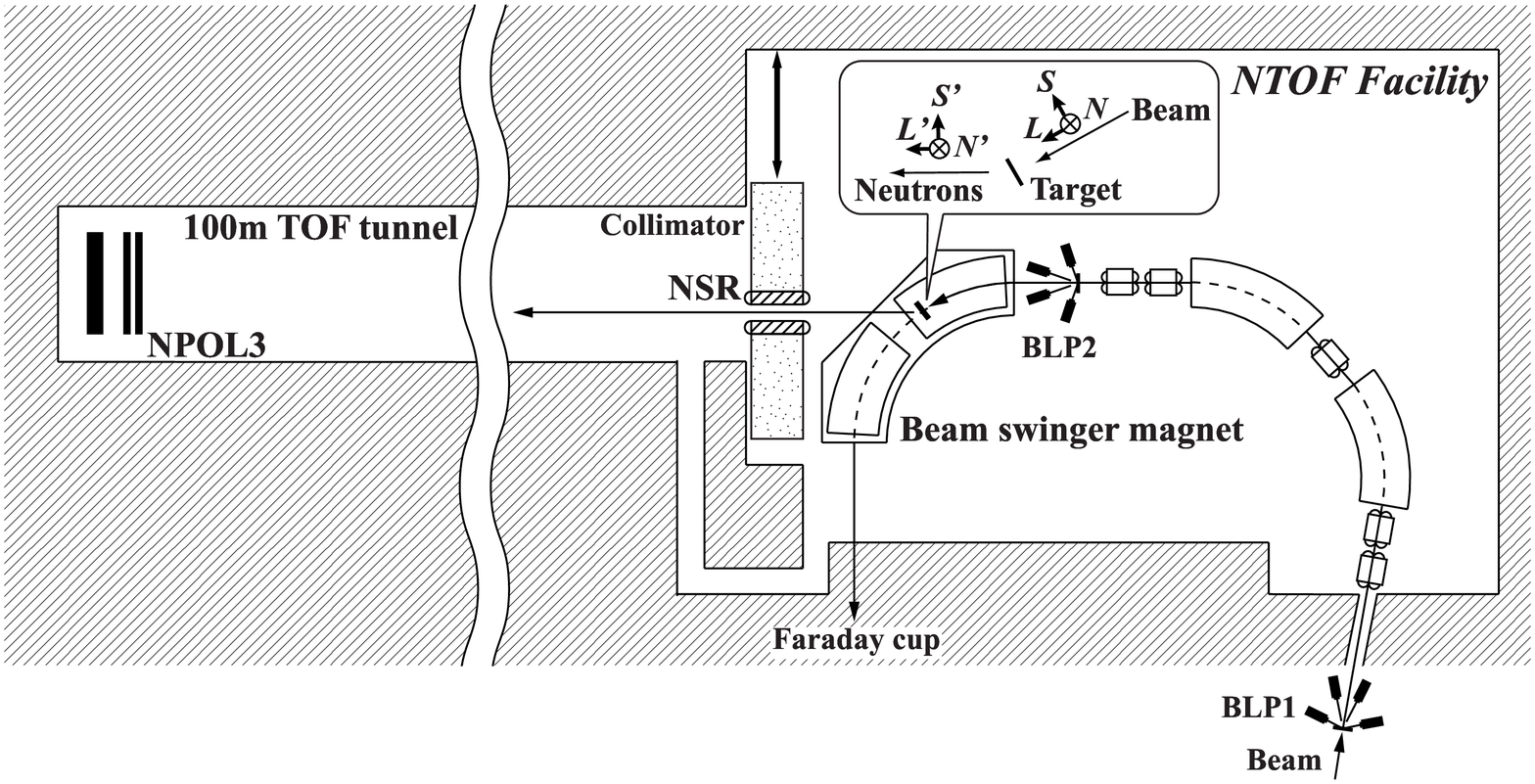}
\caption{A schematic layout of the NTOF facility (not to scale). 
The coordinate systems for incident protons and outgoing neutrons 
are also shown. 
$S$ (Sideways), $N$ (Normal) and $L$ (Longitudinal) 
form a right-handed system for incident protons and 
$S'$, $N'$, $L'$ for outgoing neutrons.} 
\label{fig:ntof}
\end{figure}

\begin{figure}[t]
\includegraphics[width=\hsize,clip]{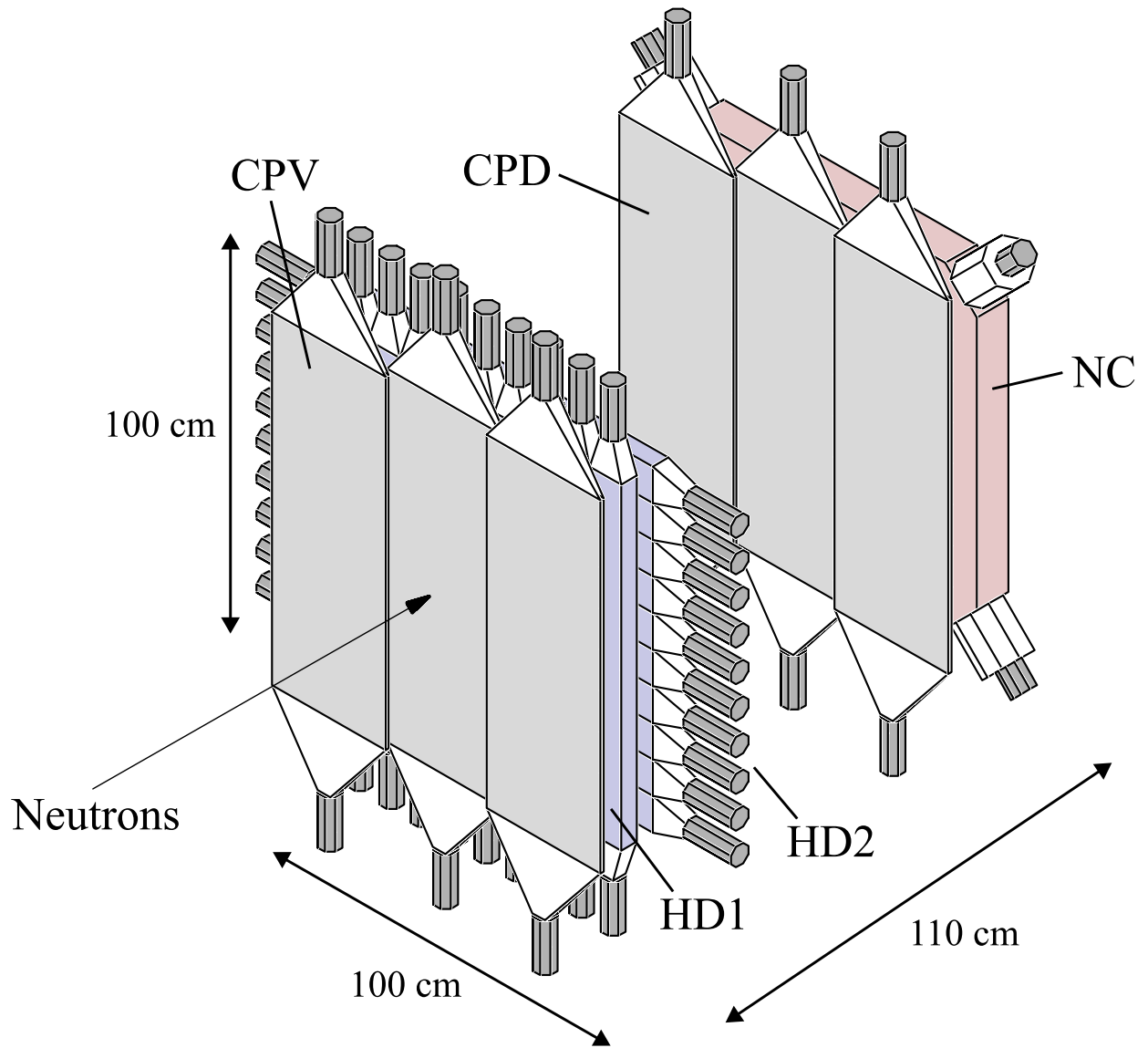}
\caption{(Color online) A schematic view of the neutron 
detector/polarimeter NPOL3. 
In the detector mode, HD1 and HD2 act as neutron detectors. 
In the polarimetry mode, HD1 and HD2 are the 
analyzer planes while NC is the catcher plane. 
Thin plastic scintillator planes are used to veto (CPV) 
or identify (CPD) charged particles.} 
\label{fig:npol3}
\end{figure}

The experiment was performed using the Neutron Time-Of-Flight (NTOF) 
facility~\cite{sakai1996} and the neutron detector/polarimeter 
NPOL3~\cite{wakasa2005-1} 
at the Research Center for Nuclear Physics (RCNP) at Osaka University. 
Detailed descriptions of the NTOF facility and the NPOL3 system 
can be found in 
Refs.~\cite{sakai1996,wakasa1998,wakasa2005-1,wakasa2007-1,wakasa2007-2,dozono2008,wakasa2008,ihara2008}. 
Only the details relevant to the 
present experiment are described here. 
Schematic layouts of the NTOF facility and the NPOL3 system 
are presented in Figs.~\ref{fig:ntof} and~\ref{fig:npol3}, respectively. 
In Fig.~\ref{fig:ntof},  
the coordinate systems for incident protons and outgoing neutrons 
are also shown.

\subsection{\label{polarized_proton_beam}Polarized proton beam}
The high intensity polarized ion source (HIPIS)~\cite{hatanaka1997} 
at RCNP was used to provide the proton beam. 
The direction of the beam polarization was reversed every 5 s 
in order to minimize geometrical false asymmetries 
that might be present in the experimental apparatus. 
The polarized proton beam from HIPIS was 
accelerated up to $T_p=53$ and 296~MeV by 
the Azimuthally Varying Field (AVF)~\cite{kondo1975} 
and Ring~\cite{miura1993} cyclotrons, respectively. 
The radio frequency (rf) of the AVF cyclotron was 15.42~MHz, 
corresponding to a beam pulse period of 64.86~ns. 
In the cross section and analyzing power measurement, 
one out of every five beam pulses 
was selected prior to injection into the Ring cyclotron, 
yielding a pulse period of 324.3~ns. 
This pulse selection reduced the wrap-around events of slow neutrons 
from preceding beam pulses. 
In the measurement of the polarization transfer observables, 
the pulse selection was not performed, so as 
to achieve reasonable statistical accuracy. 
Note that the contribution from the wrap-around events was 
negligibly small ($\lesssim 1\%$) because of the kinematical 
selection at the NPOL3 system 
(see Sec.~\ref{sec:effective_analyzing_power}). 
The single-turn extraction was maintained during these measurements 
in order to preserve the beam polarization.

Two superconducting solenoid magnets (SOL1 and SOL2) 
located in the injection line from the AVF to Ring cyclotrons 
were used to precess the proton spin. 
Each magnet can rotate the direction of the polarization vector from 
the normal $\hat{N}$ into the sideways $\hat{S}$ directions. 
These two magnets are separated by a bending angle of $45^{\circ}$, 
and thus the spin precession angle in this bending magnet is about 
$85.2^{\circ}$ for $T_p=53~{\rm MeV}$ protons. 
Therefore, proton beams are obtained with longitudinal ($\hat{L}$) 
and sideways ($\hat{S}$) polarizations at the exit of SOL2 
by using the SOL1 and SOL2 magnets, respectively. 

The beam polarization was continuously 
monitored by two sets of beam line polarimeters 
(BLP1 and BLP2)~\cite{sakai1996,wakasa1998} 
after the Ring cyclotron. 
Each polarimeter consists of four conjugate-angle pairs of 
plastic scintillation detectors and 
determines the beam polarization via 
$^{1}{\rm H}(\vec{p},p)^{1}{\rm H}$ scattering 
in the normal ($\hat{N}$) and sideways ($\hat{S}$) directions. 
A self-supporting ${\rm CH}_2$ target with 
a thickness of $1.1~{\rm mg/cm}^2$ was used as the hydrogen target, 
and the elastically scattered and recoil protons were detected 
in coincidence with a conjugate-angle pair of plastic scintillators. 
BLP1 and BLP2 are separated by a bending angle of 
$98^{\circ}$, and thus the spin precession angle in the bending 
magnet is about $231.1^{\circ}$ for $T_p=296~{\rm MeV}$ protons. 
Therefore, all components 
($p_S$, $p_N$, $p_L$) of the beam polarization can be simultaneously determined 
using BLP1 and BLP2. 
The typical magnitude of the beam polarization was about 0.70. 

\subsection{\label{sec:NTOF_facility}Target and NTOF facility}
The NTOF facility~\cite{sakai1996}, as illustrated in
Fig.~\ref{fig:ntof}, 
consists of a beam swinger magnet, 
a neutron spin rotation (NSR) magnet, 
and a 100~m Time-Of-Flight (TOF) tunnel. 
The proton beam bombarded a self-supporting 
$^{\rm nat}{\rm C}$ (98.9\% $^{12}{\rm C}$) target with a thickness 
of 89~${\rm mg/cm}^2$ in the beam swinger magnet. 
The target energy loss was estimated to be about 270~keV 
by using the stopping power of carbon for 296~MeV protons. 
Neutrons from the target entered the TOF tunnel and were detected 
using NPOL3 at the end of its flight path. 
Protons downstream of the target were swept up
by the beam swinger magnet 
and stopped by a graphite beam stop (Faraday cup) 
from which the integrated beam current was measured. 
Typical beam currents were 50 and 500~nA for the cross section 
and polarization transfer measurements, respectively. 
The reaction angle was changed by repositioning the target along 
the beam trajectory inside the beam swinger magnet. 

The NSR magnet was positioned at the entrance of 
the TOF tunnel. 
This magnet was used to precess the neutron polarization from 
the longitudinal $\hat{L}'$ to the normal $\hat{N}'$ directions, 
enabling the measurement of the longitudinal component of 
the neutron polarization with NPOL3 as the normal component. 
This magnet was also used for the measurement of 
the induced polarization $P$~\cite{wakasa1999}. 
In this case, the neutron polarization was precessed 
in the $N'$ -- $L'$ plane by about $120^{\circ}$, 
depending on the neutron kinetic energy. 

\subsection{\label{sec:npol3}Neutron detector/polarimeter NPOL3}
The NPOL3 system~\cite{wakasa2005-1}, 
illustrated in Fig.~\ref{fig:npol3}, 
consists of three planes of scintillation detectors. 
Each of the first two planes (HD1 and HD2) consists of 
10~sets of one-dimensional 
position-sensitive plastic scintillation counters (hodoscopes) 
with a size of $100 \times 10 \times 5~{\rm cm}^3$, covering an area of $100 \times 100~{\rm cm}^2$. 
The last plane (NC) is a two-dimensional 
position-sensitive liquid scintillation counter with 
a size of $100 \times 100 \times 10~{\rm cm}^3$. 
Both HD1 and HD2 served as neutron detectors and 
neutron polarization analyzers for the cross section 
and polarization transfer measurements, respectively, 
and NC acted as a catcher for the particles scattered 
by HD1 or HD2 in the polarization transfer measurements. 
Thin plastic scintillator planes (CPV and CPD) in front of HD1 and NC 
were used to veto and identify charged particles, respectively. 

The neutron energy was determined by the TOF to a given hodoscope 
with respect to the rf signal of the AVF cyclotron. 
A peak from $^{12}{\rm C}(p,n)^{12}{\rm N}({\rm g.s.})$ provided 
a time reference for the absolute timing calibration. 
The overall energy resolution in full width at half maximum 
(FWHM) was about 500~keV, 
mainly due to the target energy loss and the beam pulse width 
with contributions of about 270 and 350~keV, respectively. 

\section{\label{sec:data_reduction}DATA REDUCTION}

\subsection{\label{sec:polarization_observables}
Polarization observables}
A complete set of polarization observables, $A_y$, $P$, and 
$D_{ij}(i=S',N',L'; j=S,N,L)$, 
for a parity conserving reaction 
relates the outgoing neutron polarization 
$\mbox{\boldmath $p$}'=(p_{S'}',p_{N'}',p_{L'}')$ to the 
incident proton polarization 
$\mbox{\boldmath $p$}=(p_{S},p_{N},p_{L})$ according to 
\begin{eqnarray}
\left(
\begin{array}{c}
p_{S'}' \\
p_{N'}' \\
p_{L'}'
\end{array}
\right)& = &
\left[
\left(
\begin{array}{ccc}
D_{S'S} & 0 & D_{S'L} \\
0 & D_{NN} & 0 \\
D_{L'S} & 0 & D_{L'L}
\end{array}
\right)
\left(
\begin{array}{c}
p_{S} \\
p_{N} \\
p_{L}
\end{array}
\right)
\right. \nonumber \\
& & +\left.
\left(
\begin{array}{c}
0 \\
P \\
0
\end{array}
\right)
\right]
{\displaystyle \frac{1}{1+p_{N}A_y}}.
\label{eq:polarization_observable}
\end{eqnarray}
The directions of the coordinate system 
(sideways $S$, normal $N$, and longitudinal $L$) are 
defined in terms of the incident proton momentum 
$\mbox{\boldmath $k$}_{\rm lab}$
and the outgoing neutron momentum $\mbox{\boldmath $k'$}_{\rm lab}$ 
in the laboratory frame as 
$\mbox{\boldmath $\hat{L}$}=\mbox{\boldmath $\hat{k}$}_{\rm lab}$, 
$\mbox{\boldmath $\hat{L'}$}=\mbox{\boldmath $\hat{k'}$}_{\rm lab}$, 
$\mbox{\boldmath $\hat{N}$}=\mbox{\boldmath $\hat{N'}$}=
(\mbox{\boldmath $\hat{k}$}_{\rm lab} \times 
\mbox{\boldmath $\hat{k'}$}_{\rm lab})
/|\mbox{\boldmath $\hat{k}$}_{\rm lab} \times 
\mbox{\boldmath $\hat{k'}$}_{\rm lab}|$, 
$\mbox{\boldmath $\hat{S}$}=\mbox{\boldmath $\hat{N}$}
\times \mbox{\boldmath $\hat{L}$}$, and 
$\mbox{\boldmath $\hat{S'}$}=\mbox{\boldmath $\hat{N'}$}
\times \mbox{\boldmath $\hat{L'}$}$. 

The analyzing power $A_y$, the induced polarization $P$, 
and the polarization transfer observable $D_{NN}$ were measured for a normally ($\hat{N}$) polarized proton beam. 
The other polarization transfer observables, $D_{S'S}$, 
$D_{L'S}$, $D_{S'L}$, and $D_{L'L}$, were obtained from 
measurements with two kinds of proton beams 
polarized in the horizontal plane. 
Note that the polarization axes of these beams were 
almost orthogonal to each other. 
Therefore, the efficiency of measuring $D_{ij}$ 
is almost the same as that for pure sideways ($\hat{S}$) and 
longitudinal ($\hat{L}$) polarized proton beams~\cite{wakasa1999}. 

\subsection{\label{sec:neutron_detection_efficiency}
Neutron detection efficiency}

The differential cross section
 $(d\sigma/d\Omega)_{\rm lab}$ 
in the laboratory frame is related to 
the observed neutron yield $N_n$ as 
\begin{equation}
\left(\frac{d\sigma}{d\Omega}\right)_{\rm lab}
=\frac{N_n}{N_p \rho \Delta \Omega \varepsilon T f_{\rm live}}, 
\label{eq:cross_section}
\end{equation}
where $N_p$ is the number of incident protons, 
$\rho$ is the target density, $\Delta \Omega$ and 
$\varepsilon$ are the solid angle and 
intrinsic neutron detection efficiency of NPOL3 (HD1 and HD2), 
respectively, 
$T$ is the neutron transmission factor along the flight path in the air, 
and $f_{\rm live}$ is the live time ratio. 

The product $\varepsilon T$ was determined by measuring 
the neutron yield from the 
$0^{\circ}$ $^{7}{\rm Li}(p,n)^{7}{\rm Be}({\rm g.s.}+0.43~{\rm MeV})$ 
reaction which has a constant center of mass (c.m.) cross section of 
$(d\sigma/d\Omega)_{\rm c.m.}=27.0 \pm 0.8~{\rm mb/sr}$ 
at an incident energy 
range of $T_p=80$--795~MeV~\cite{taddeucci1990}. 
A self-supporting $^7{\rm Li}$ (99.97\%) target 
with a thickness of $54~{\rm mg/cm}^2$ was used. 
The $\varepsilon T$ value was $0.051 \pm 0.003$ where 
the uncertainty comes mainly from 
the uncertainties in the cross section for the 
$0^{\circ}\ ^{7}{\rm Li}(p,n)^{7}{\rm Be}({\rm g.s.}+ 0.43~{\rm MeV})$ 
reaction (3\%) and in the thickness of the $^7{\rm Li}$ target (3\%). 
We note that the transmission factor $T$ has been assumed to 
be independent of reaction angle because the dependence of 
the flight length on reaction angle is very small ($\lesssim 1\%$).

\subsection{\label{sec:effective_analyzing_power}
Effective analyzing power}
The neutron polarization was analyzed using 
$\vec{n}+p$ and quasi-elastic $\vec{n}+{\rm C}$ scattering 
in analyzer HD1 or HD2, and the recoiled protons 
were detected with catcher NC. 
These events were kinematically resolved from background events 
such as wrap-around and the target gamma rays 
by using time and position information from the analyzer and catcher planes. 
Both the normal $N'$ and sideways $S'$ components of the neutron 
polarization were measured simultaneously, with 
an azimuthal distribution of $\vec{n}+p$ and 
quasi-elastic $\vec{n}+{\rm C}$ events. 
 
The effective analyzing power $A_{y;{\rm eff}}$ of NPOL3 
was measured using polarized neutrons from 
the GT $^{12}{\rm C}(\vec{p},\vec{n})^{12}{\rm N}({\rm g.s.})$ reaction 
at $\theta_{\rm lab}=0^{\circ}$. 
Two kinds of polarized protons having normal ($p_{N}$) and 
longitudinal ($p_{L}$) polarizations were used. 
The corresponding neutron polarizations are 
$p'_{N}=p_{N}D_{NN}(0^{\circ})$ and 
$p'_{L}=p_{L}D_{LL}(0^{\circ})$ 
where $D_{NN}(0^{\circ})$ and $D_{LL}(0^{\circ})$ represent 
the polarization transfer observables at $\theta_{\rm lab}=0^{\circ}$. 
Then the asymmetries, $\epsilon_N$ and $\epsilon_L$, measured by NPOL3 are 
\begin{equation}
\begin{array}{lcccl}
\epsilon_N & = & p'_{N} A_{y;{\rm eff}} & = & 
 p_N D_{NN}(0^{\circ}) A_{y;{\rm eff}}, \\
\epsilon_L & = & p'_{L} A_{y;{\rm eff}} & = & 
p_{L}D_{LL}(0^{\circ}) A_{y;{\rm eff}}. 
\label{eq:Ayeff_method2}
\end{array}
\end{equation}
As described in Sec~\ref{sec:NTOF_facility}, 
the asymmetry $\epsilon_L$ was measured 
as the normal component using the NSR magnet. 
Because the polarization transfer observables 
$D_{ii}(0^{\circ})$ for the GT transition 
satisfy~\cite{wakasa2004-1}
\begin{equation}
2D_{NN}(0^{\circ})+D_{LL}(0^{\circ})=-1, 
\label{eq:Ayeff_method1} 
\end{equation} 
$A_{y{\rm ;eff}}$ can be expressed as 
\begin{equation}
A_{y;{\rm eff}}=
-\left(
2 \frac{\epsilon_N}{p_N}
+ \frac{\epsilon_L}{p_L}
\right)
\label{eq:Ayeff_method3}
\end{equation}
using Eqs.~(\ref{eq:Ayeff_method2}) and (\ref{eq:Ayeff_method1}). 
Thus, one obtains the $A_{y{\rm ;eff}}$ value without 
knowing ahead of time the $D_{ii}(0^{\circ})$ values. 
The resulting $A_{y{\rm ;eff}}$ is 
$0.191 \pm 0.016$ 
where the uncertainty includes 
the statistical ($\simeq 6\%$) and 
systematic ($\simeq 2\%$) uncertainties. 
The systematic uncertainty is estimated by 
considering the uncertainty of the beam polarization~\cite{wakasa1998}. 

\subsection{\label{sec:peak_fitting}Peak fitting}

\begin{figure}[t]
\includegraphics[width=0.9\hsize,clip]{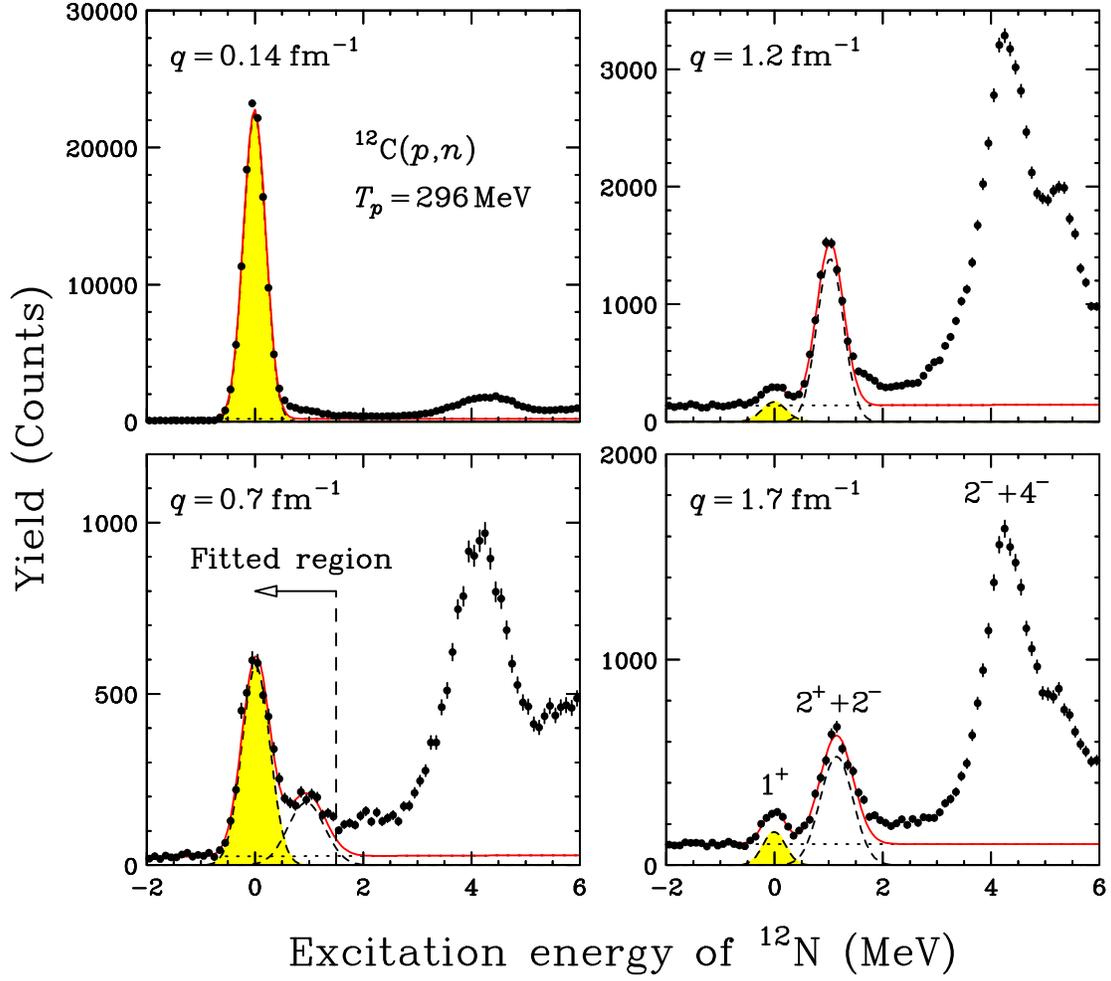}
\caption{(Color online) Excitation energy spectra for 
$^{12}{\rm C}(p,n)^{12}{\rm N}$ at $T_p=296~{\rm MeV}$ and 
$q=0.14$, 0.7, 1.2, and $1.7~{\rm fm}^{-1}$. 
The dashed curves are fits to the individual peaks. 
The solid curves indicate the sum of the peak contributions 
including the background plotted as the straight dotted lines.} 
\label{fig:spectrum}
\end{figure}

Figure~\ref{fig:spectrum} shows typical excitation energy spectra 
of $^{12}{\rm C}(p,n)^{12}{\rm N}$ at four momentum transfers of 
$q=0.14$, 0.7, 1.2, and $1.7~{\rm fm}^{-1}$. 
The GT $1^+$ state at $E_x=0~{\rm MeV}$ (ground state of $^{12}{\rm N}$) 
gives rise to a prominent peak at small momentum transfers. 
At large momentum transfers, on the other hand, 
its peak is small and not fully resolved from 
a large peak consisting of 
excited states with $J^{\pi}=2^+$ and $2^-$ 
at $E_x=0.96$ and $1.19~{\rm MeV}$, respectively. 
Therefore, peak fitting was performed to extract the 
yield of the GT $1^+$ state. 
The spectra were fitted at $E_x<1.5~{\rm MeV}$ 
where the excited states at $E_x=0.96$ and $1.19~{\rm MeV}$ 
were treated as a single peak 
because the present energy resolution could not resolve them. 
The continuum background from wrap-around and $^{13}{\rm C}(p,n)$ 
events was considered to be a linear function of $E_x$. 
The dashed curves in Fig.~\ref{fig:spectrum} show the 
fitting results for the individual peaks, while the solid curves 
show the sum of these contributions including the background 
indicated as the straight dotted lines. 
The peak fittings at all momentum transfers sufficed to extract the GT $1^+$ yield. 

\section{\label{sec:results}RESULTS AND ANALYSIS}
\subsection{\label{sec:cross_section}
Cross section and polarization observables}

Figure~\ref{fig:I_result} shows the cross section for the 
$^{12}{\rm C}(p,n)^{12}{\rm N}({\rm g.s.},1^+)$ reaction 
at $T_p=296~{\rm MeV}$ as a function of the momentum transfer $q$. 
The corresponding reaction angle $\theta_{\rm c.m.}$ 
is also shown on the top of the figure. 
The momentum transfer resolution is about $0.04~{\rm fm}^{-1}$ 
which is mainly due to the finite solid angle of the detector. 
As seen in Fig.~\ref{fig:spectrum}, 
the GT state is not clearly resolved from the neighboring states, 
and thus there is a correlation between the yields of these 
two components in the peak fitting. 
By considering the uncertainties of the GT yields 
due to this correlation, 
we have estimated the systematic uncertainties. 
The shaded boxes show the uncertainties including 
statistical and systematic uncertainties. 
The statistical and systematic uncertainties 
at large momentum transfers of $q \simeq 1.6~{\rm fm}^{-1}$ 
are about 2\% and 4\%, respectively. 
A 6\% uncertainty due to the cross section normalization 
(see Sec.~\ref{sec:neutron_detection_efficiency})
is not included. 
The open circles and open triangles, respectively, are data 
for the same reaction~\cite{wakasa1995} 
and the analogous $^{12}{\rm C}(p,p')^{12}{\rm C}^*(1^+,T=1)$ 
scattering~\cite{tamii2007} at $T_p=295~{\rm MeV}$. 
The $^{12}{\rm C}(p,p')$ data have been multiplied by 
a factor of two because of the difference in 
the isospin Clebsch-Gordan (CG) coefficients between  
$^{12}{\rm C}(p,n)$ and $^{12}{\rm C}(p,p')$. 
Our data are consistent with the previous data within 
the statistical and systematic uncertainties. 

\begin{figure}[t]
\includegraphics[width=0.8\hsize,clip]{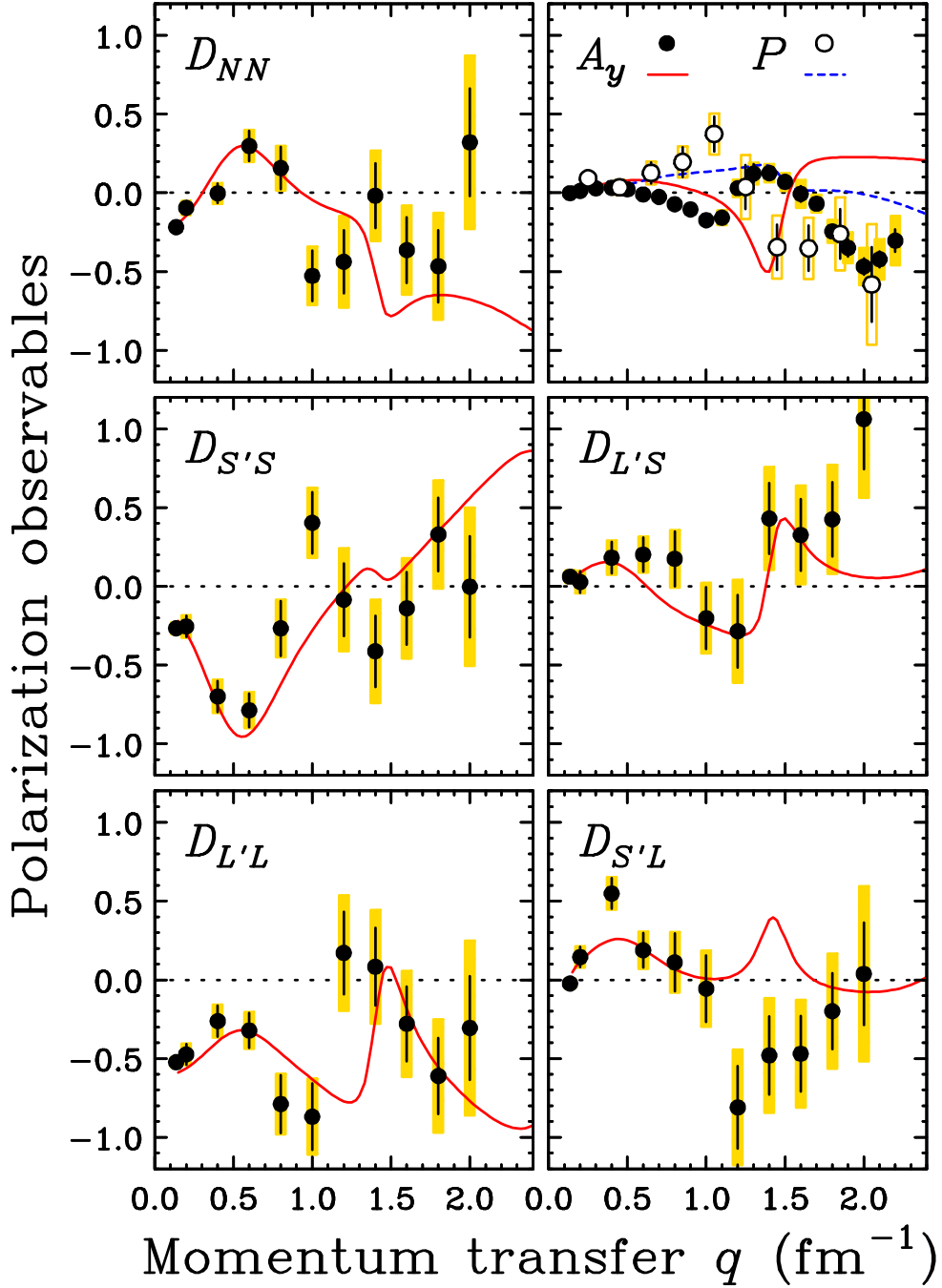}
\caption{(Color online) Measured polarization transfer observables 
$D_{ij}$, analyzing power $A_y$, and induced polarization $P$ 
for $^{12}{\rm C}(p,n)^{12}{\rm N}({\rm g.s.},1^+)$ at 
$T_p=296~{\rm MeV}$. 
The induced polarization data $P$ 
are offset by a momentum transfer of $0.05~{\rm fm}^{-1}$ 
so that the analyzing power $A_y$ and induced polarization $P$ 
data can be distinguished. 
The solid and dashed curves are the results of DWIA calculations 
with a shell-model wave function.} 
\label{fig:Dij_result}
\end{figure}

In Fig.~\ref{fig:Dij_result}, a complete set of polarization 
observables, $D_{ij}$, $A_y$, and $P$, are presented for the 
$^{12}{\rm C}(\vec{p},\vec{n})^{12}{\rm N}({\rm g.s.},1^+)$ reaction 
at $T_p=296~{\rm MeV}$ as a function of momentum transfer. 
In the top right panel, the $A_y$ and $P$ data are shown 
as filled and open circles, respectively, 
and the $P$ data are offset by a momentum transfer 
of $0.05~{\rm fm}^{-1}$ 
for clarity. 
The error bars represent statistical uncertainties only, 
while the shaded and open boxes include 
the systematic uncertainties. 
The statistical and systematic uncertainties in $D_{ij}$ 
for large momentum transfers of $q \gtrsim 1.0~{\rm fm}^{-1}$ 
are about 0.19--0.24 and 0.21--0.26, respectively, 
which are satisfactory for discussing nuclear correlation effects 
in this momentum transfer range. 

\subsection{\label{sec:DWIA_calculations}
DWIA calculations with shell-model wave function}
We performed microscopic DWIA calculations using the computer code 
{\sc dw81}~\cite{dw81}, 
which treats the knock-on exchange amplitude exactly. 
Distorted waves were generated using a global 
optical model potential (OMP) optimized for $^{12}{\rm C}$ 
in the proton energy range of 
$T_p=20$--$1040~{\rm MeV}$~\cite{hama1990,cooper1993}, 
with the Coulomb term turned off for the exit channel. 
The nucleon-nucleon ($NN$) $t$-matrix parameterized 
by Franey and Love (FL)~\cite{franey1985} at 325~MeV 
was used as the interaction between the incident and struck nucleons. 
The one-body density matrix elements (OBDMEs) were 
obtained from shell-model calculations using 
the computer code {\sc oxbash}~\cite{oxbash}. 
These calculations were performed 
in the $0\hbar \omega$ $p$-shell model space 
using the Cohen-Kurath wave functions 
(CKWFs)~\cite{cohen1965} based on the (6--16) 2BME interaction. 
The transition form factor was normalized to reproduce the observed 
beta-decay $ft$ value of $13178~{\rm s}$~\cite{alburger1978} 
which corresponds to a GT strength $B({\rm GT})$ of
0.873~\cite{schreckenbach1995}. 
The radial part of the single-particle wave functions was 
generated by a Woods-Saxon (WS) potential 
with $r_0=1.27~{\rm fm}$, $a_0=0.67~{\rm fm}$~\cite{bohr}, 
and a spin-orbit potential depth of 
$V_{\rm so}=6.5~{\rm MeV}$~\cite{nishida1995}. 
The depths of the WS potential were adjusted to 
reproduce the separation energies for the $0p_{3/2}$ orbits. 

The solid curves in 
Figs.~\ref{fig:I_result} and~\ref{fig:Dij_result} 
show the results of the calculations. 
The normalization factor for the transition form factor 
is $N=0.94$. 
These calculations reproduce the experimental data 
reasonably well at small momentum transfers of $q \lesssim 0.5~{\rm fm}^{-1}$ 
but show poor agreement with the data at $q \gtrsim 0.5~{\rm fm}^{-1}$. 
In particular, the calculations 
shift the momentum transfer dependence of the cross section 
to larger momentum transfers and underestimate the cross section 
at $q \simeq 1.6~{\rm fm}^{-1}$. 
In order to investigate the reason for this discrepancy, 
we next separated the cross section into 
polarized cross sections using the polarization observables. 

\subsection{\label{sec:polarized_cross_section}
Polarized cross sections}

\begin{figure}[t]
\includegraphics[width=0.9\hsize,clip]{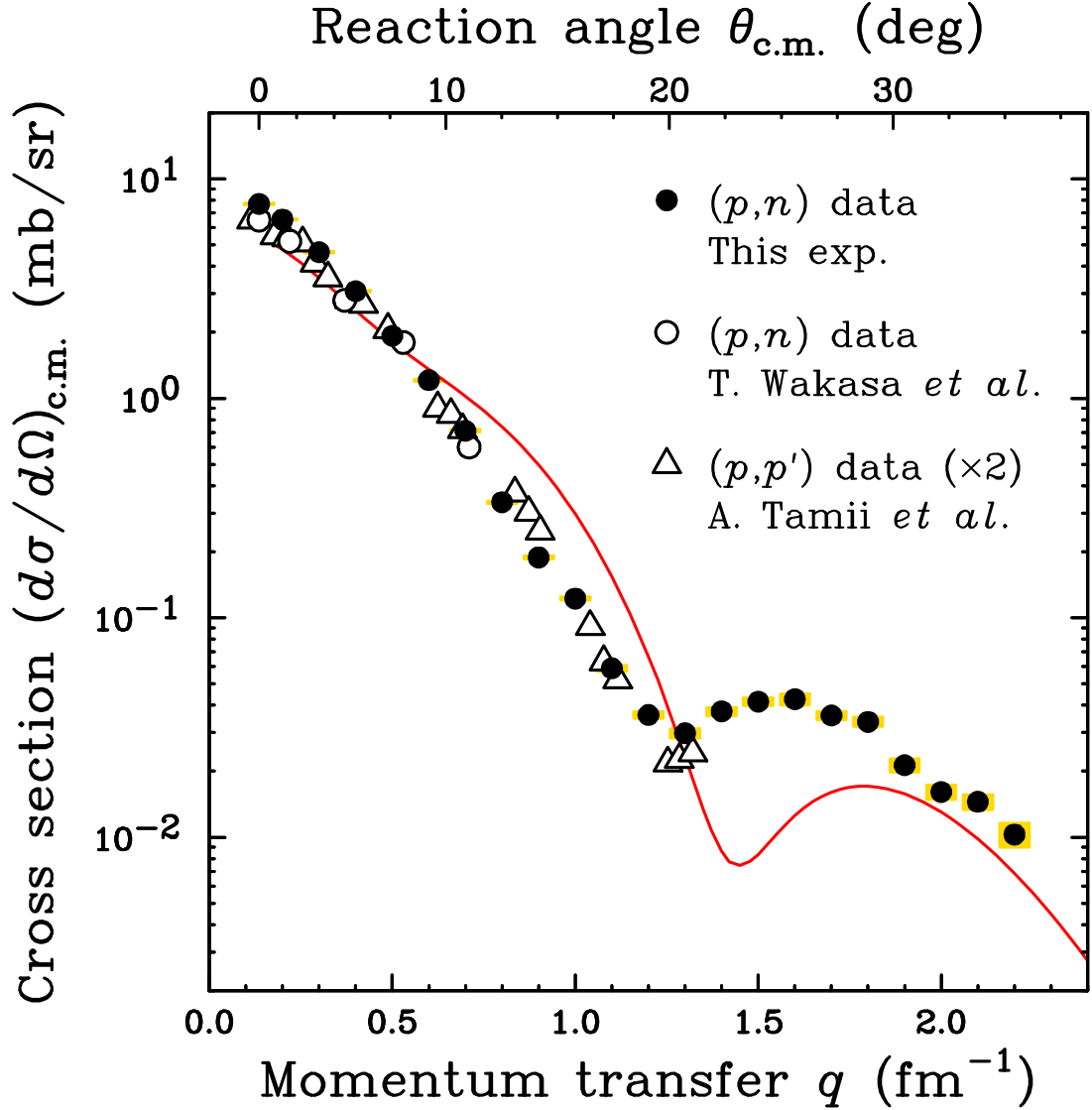}
\caption{(Color online) Measured cross sections for 
$^{12}{\rm C}(p,n)^{12}{\rm N}({\rm g.s.},1^+)$ at 
$T_p=296~{\rm MeV}$ (filled circles) 
as a function of momentum transfer. 
The corresponding reaction angle $\theta_{\rm c.m.}$ is 
also shown on the top of the figure. 
The open circles are data at $T_p=295~{\rm MeV}$~\cite{wakasa1995}. 
The open triangles are 
$^{12}{\rm C}(p,p')^{12}{\rm C}^*(1^+,T=1)$ data at 
$T_p=295~{\rm MeV}$~\cite{tamii2007}, multiplied by a factor of two 
as described in the text. 
The solid curve shows DWIA calculations 
using a shell-model wave function.} 
\label{fig:I_result}
\end{figure}

\begin{figure}[t]
\includegraphics[width=0.9\hsize,clip]{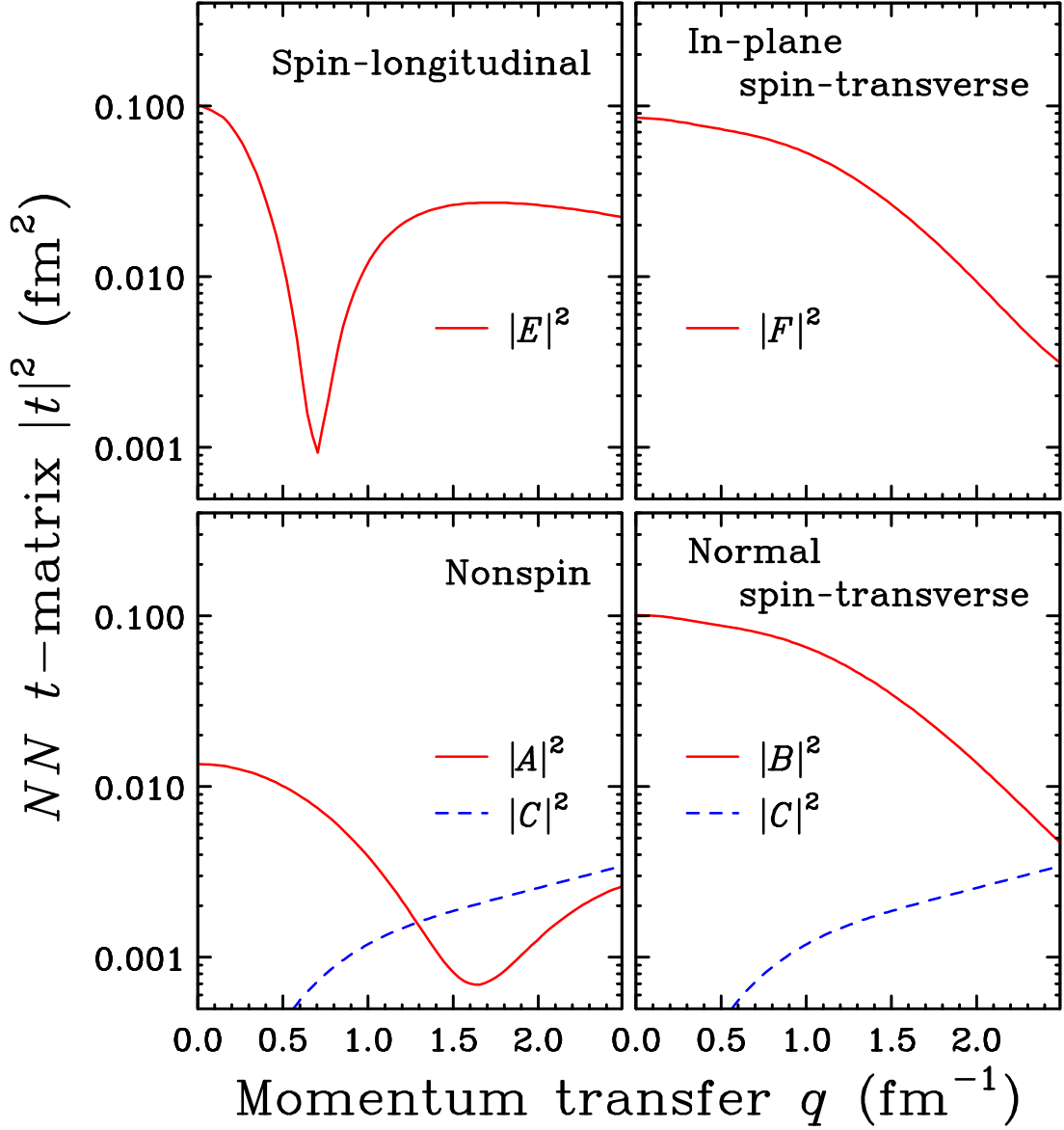}
\caption{(Color online) Squared $t$-matrix components 
calculated from the FL $t$-matrix at 325~MeV. 
} 
\label{fig:t-matrix}
\end{figure}

The cross section $I$ 
($(d\sigma/d\Omega)_{\rm c.m.}$ in Fig.~\ref{fig:I_result}) 
can be separated into four polarized cross sections $ID_i$ as 
\begin{equation}
I=ID_0 + ID_q + ID_n + ID_p,  
\label{eq:IDi}
\end{equation}
where $D_i$ are the c.m. polarization observables 
introduced by Bleszynski {\it et al.}~\cite{bleszynski1982}. 
The c.m. coordinate system $(q,n,p)$ is defined as 
$\hat{\mbox{\boldmath $q$}}=
(\mbox{\boldmath $k'$}-\mbox{\boldmath $k$})/
(|\mbox{\boldmath $k'$}-\mbox{\boldmath $k$}|)$, 
$\hat{\mbox{\boldmath $n$}}=
(\mbox{\boldmath $k$} \times \mbox{\boldmath $k'$})/
(|\mbox{\boldmath $k$} \times \mbox{\boldmath $k'$}|)$, and 
$\hat{\mbox{\boldmath $p$}}=
\hat{\mbox{\boldmath $q$}} \times \hat{\mbox{\boldmath $n$}}$, 
where $\mbox{\boldmath $k$}$ and $\mbox{\boldmath $k'$}$ 
are the momenta of the incident and outgoing nucleons in 
the c.m. frame, respectively. 
The $D_i$ values are related to $D_{ij}$ in the laboratory frame 
according to~\cite{ichimura1992}  
\begin{equation}
\begin{array}{ll}
D_0= & {\displaystyle \frac{1}{4}}
[1+D_{NN}+(D_{S'S}+D_{L'L})\cos \alpha_1 \\[10pt]
& + (D_{L'S}-D_{S'L})\sin \alpha_1],\\[15pt]
D_n= & {\displaystyle \frac{1}{4}}
[1+D_{NN}-(D_{S'S}+D_{L'L})\cos \alpha_1 \\[10pt]
& - (D_{L'S}-D_{S'L})\sin \alpha_1],\\[15pt]
D_q= & {\displaystyle \frac{1}{4}}
[1-D_{NN}+(D_{S'S}-D_{L'L})\cos \alpha_2 \\[10pt]
& - (D_{L'S}+D_{S'L})\sin \alpha_2],\\[15pt]
D_p= & {\displaystyle \frac{1}{4}}
[1-D_{NN}-(D_{S'S}-D_{L'L})\cos \alpha_2 \\[10pt]
& + (D_{L'S}+D_{S'L})\sin \alpha_2], 
\end{array}
\end{equation}
where $\alpha_1 \equiv \theta_{\rm lab} + \Omega$ and 
$\alpha_2 \equiv 2 \theta_p - \theta_{\rm lab} - \Omega$. 
Here $\theta_p$ is the angle between 
$\hat{\mbox{\boldmath $k$}}$ and $\hat{\mbox{\boldmath $p$}}$, 
and $\Omega$ is the relativistic spin rotation angle defined 
in Ref.~\cite{ichimura1992}. 
For a plane-wave impulse approximation with eikonal approximation, 
the polarized cross sections $ID_i$ can be expressed as~\cite{ichimura1992} 
\begin{equation}
\begin{array}{ll}
ID_0= & 4 K N_D 
\left(|A|^2 R_0 + |C|^2 R_n \right),\\[10pt]
ID_n= & 4 K N_D 
\left(|B|^2 R_n + |C|^2 R_0 \right),\\[10pt]
ID_q= & 4 K N_D 
\left(|E|^2 R_q + |D|^2 R_p \right),\\[10pt]
ID_p= & 4 K N_D 
\left(|F|^2 R_p + |D|^2 R_q \right), 
\end{array}
\label{eq:fia}
\end{equation}
where $K$ is a kinematical factor, $N_D$ is a distortion factor, 
$A$--$F$ are the components of the $NN$ $t$-matrix, 
and $R_i$ are the response functions. 
Figure~\ref{fig:t-matrix} shows the squared $t$-matrix components 
corresponding to each $ID_i$. 
These components are derived from the FL $t$-matrix at 325~MeV. 
The effect of the relativistic spin rotation is 
so small that the $D$ term can be neglected. 
Thus, polarized cross sections $ID_q$ and $ID_p$ 
represent spin-longitudinal ($R_q$) and spin-transverse ($R_p$) 
components exclusively. 
At forward angles, where the spin-orbit component $|C|^2$ is very small, 
polarized cross sections $ID_0$ and $ID_n$ 
represent nonspin ($R_0$) and spin-transverse ($R_n$) components, 
respectively. 

\begin{figure}[t]
\includegraphics[width=0.9\hsize,clip]{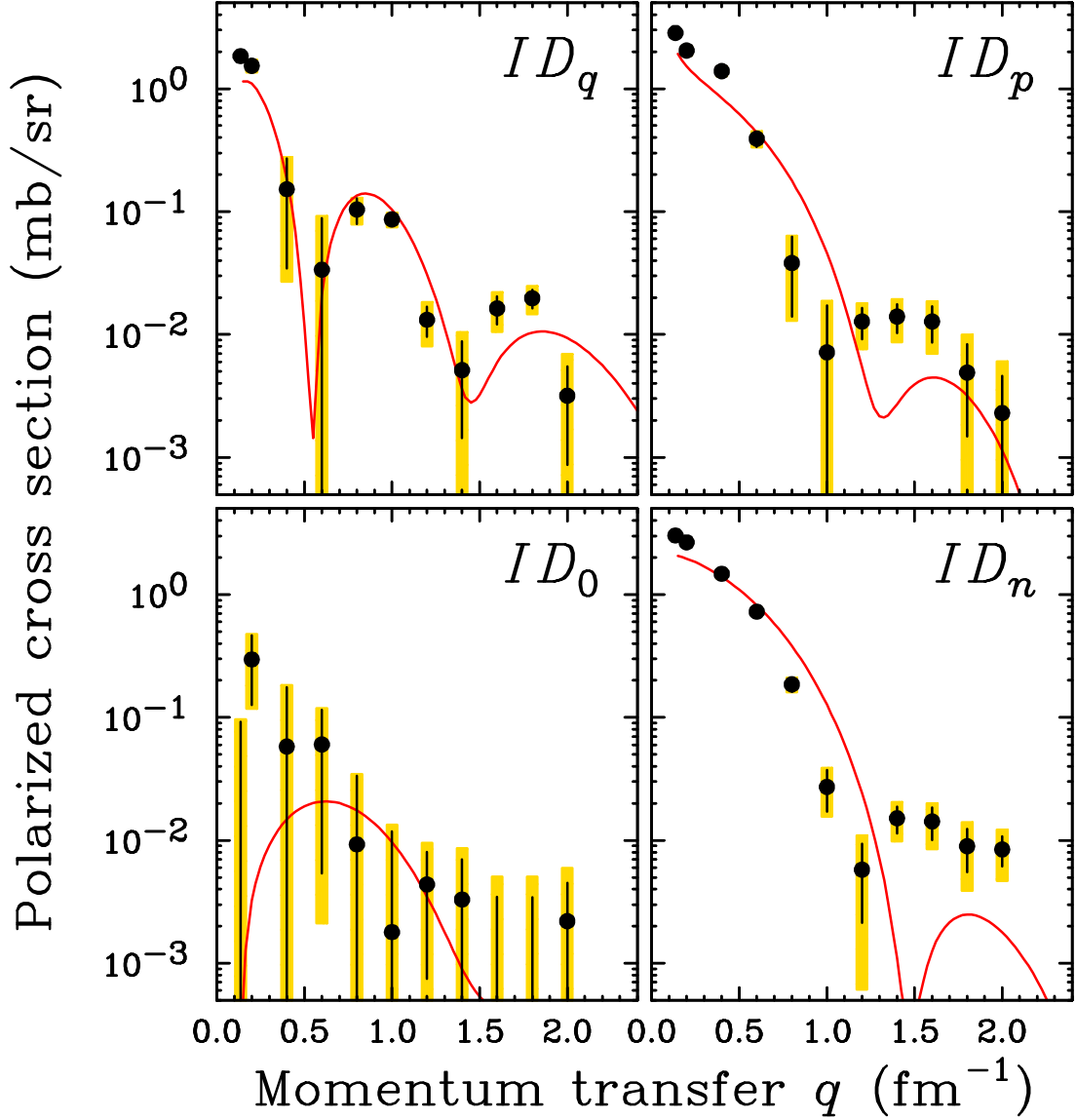}
\caption{(Color online) Measured polarized cross sections for 
$^{12}{\rm C}(p,n)^{12}{\rm N}({\rm g.s.},1^+)$ at $T_p=296~{\rm MeV}$. 
The solid curves are the results of DWIA calculations 
using the shell-model wave function. } 
\label{fig:IDi_result}
\end{figure}

Figure~\ref{fig:IDi_result} shows four polarized cross sections 
$ID_i$ as a function of momentum transfer. 
The meaning of the error bars and shaded boxes 
is the same as those in 
Figs.~\ref{fig:I_result} and \ref{fig:Dij_result}. 
Although the present GT transition does not have a
nonspin response function $R_0$, 
the nonspin polarized cross section $ID_0$ has a nonzero 
value due to the spin-orbit component $|C|^2$ 
in Eq.~(\ref{eq:fia}). 
Since $ID_0$ is small, 
we will only discuss the spin-dependent 
polarized cross sections $ID_q$, $ID_p$, and $ID_n$. 
The oscillatory pattern for the spin-longitudinal 
cross section $ID_q$ is different from those for the 
spin-transverse cross sections, $ID_p$ and $ID_n$. 
As seen in Fig.~\ref{fig:t-matrix}, 
these patterns reflect the momentum transfer dependences of 
the corresponding $NN$ $t$-matrix components. 
Compared with the spin-transverse $t$-matrix components $|B|^2$ and $|F|^2$, 
the spin-longitudinal component $|E|^2$ has the first minimum 
at lower momentum transfer of $q \simeq 0.7~{\rm fm}^{-1}$. 
This is because the real part of $E$ crosses zero 
near this momentum transfer due to the smallness of 
the pion mass. 
Thus the corresponding spin-longitudinal $ID_q$ 
shows the first minimum at $q \simeq 0.6~{\rm fm}^{-1}$. 
Therefore, the data verify suitable 
separation in the spin-longitudinal and spin-transverse modes 
based on the reaction mechanism of Ref.~\cite{ichimura1992}. 

\begin{figure}[t]
\includegraphics[width=0.8\hsize,clip]{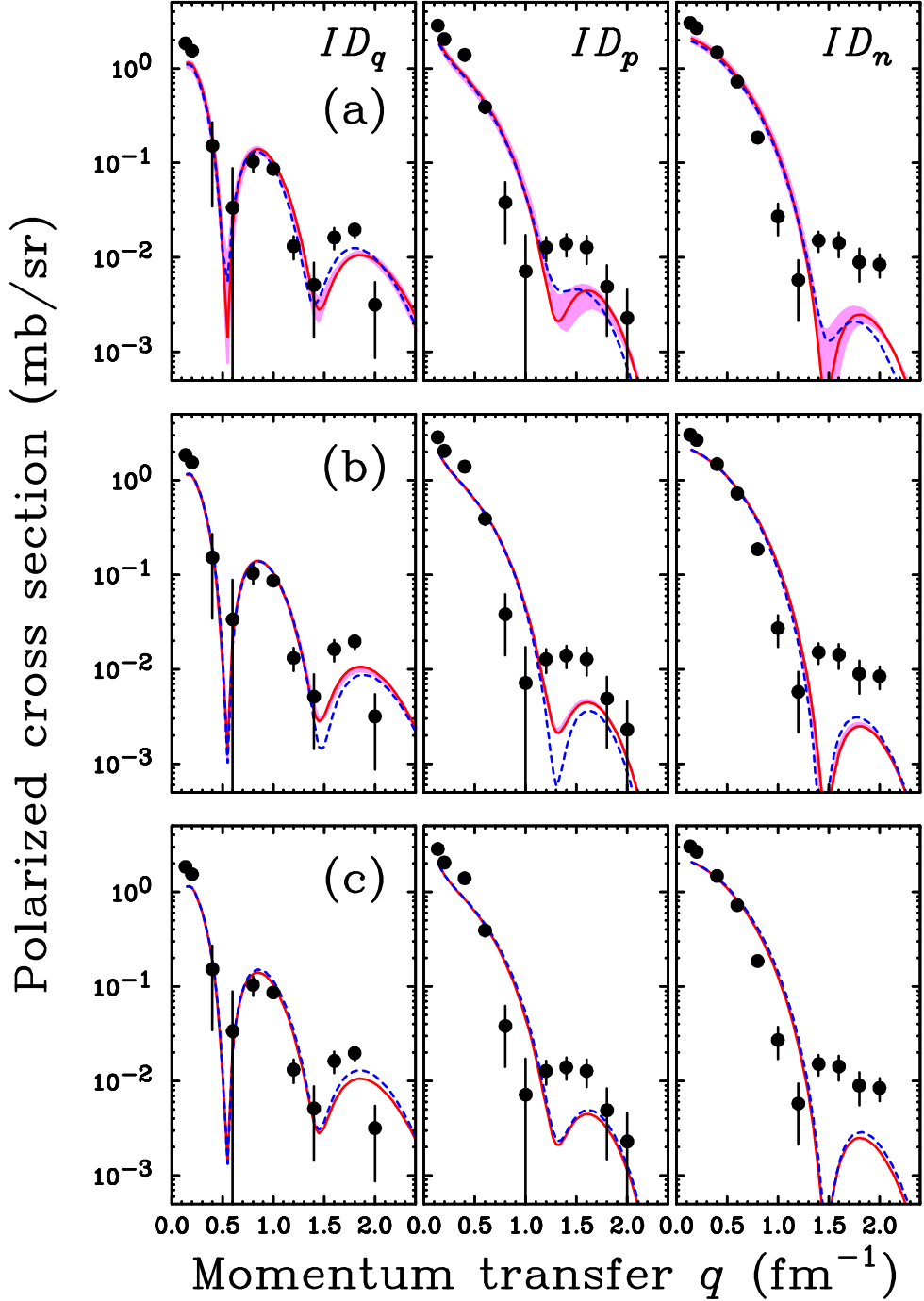}
\caption{(Color online) Parameter dependence of the calculations 
for the spin-dependent polarized cross sections $ID_q$ (left panels), 
$ID_p$ (middle panels), and $ID_n$ (right panels). 
The solid curves are the same as those in Fig.~\ref{fig:IDi_result}. 
The bands and dashed curves present DWIA results with 
other parameters: 
(a) four different OMP parameters (bands) 
and neutron global OMPs for the exit channel (dashed curves); 
(b) two different CKWFs (bands) 
and a pure $0p_{1/2}0p_{3/2}^{-1}$ configuration (dashed curves); 
and (c) the HO potential (dashed curves).} 
\label{fig:IDi_par}
\end{figure}

\begin{figure}[t]
\includegraphics[width=0.9\hsize,clip]{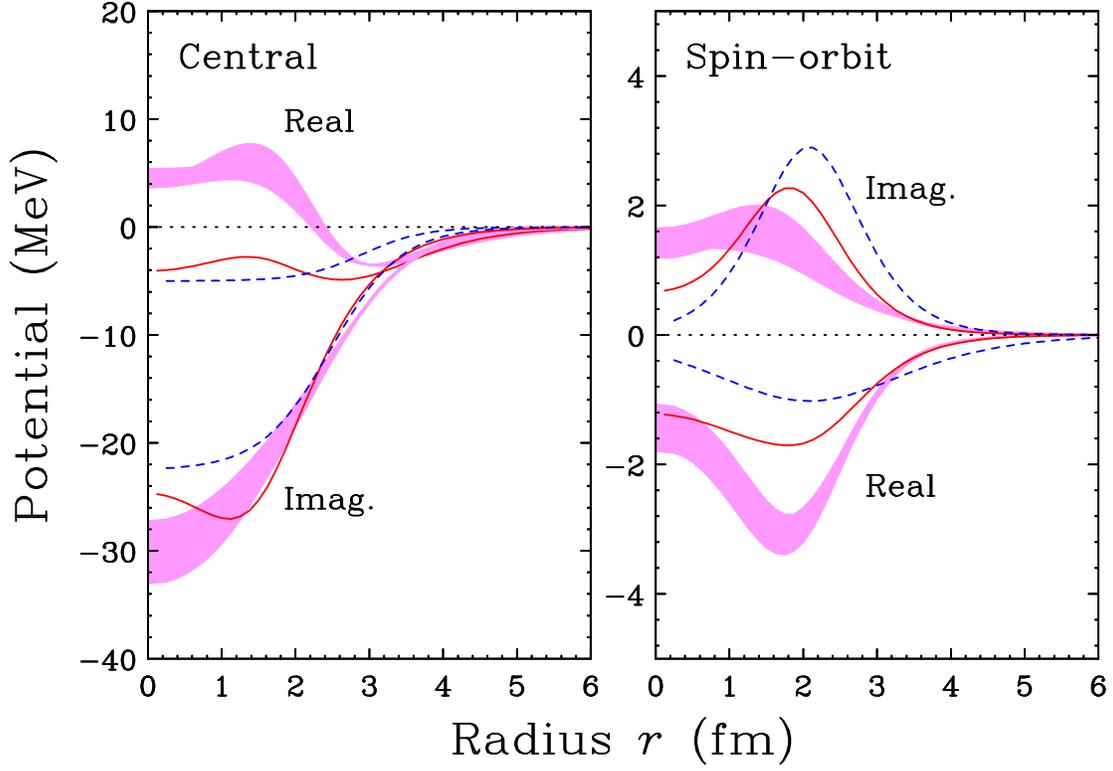}
\caption{(Color online) 
The radial dependences of the OMPs for the incident channel 
used in the DWIA calculations. 
The solid curves and the bands show the global OMPs optimized for 
$^{12}{\rm C}$ (EDAI) and $^{12}{\rm C}$--$^{208}{\rm Pb}$ (EDAI FIt 1--3) 
in the proton energy range of $T_p=20$--$1040~{\rm MeV}$, 
respectively~\cite{hama1990,cooper1993}. 
The dashed curves represent the OMP obtained from proton elastic 
scattering data on $^{12}{\rm C}$ at 
$T_p=318~{\rm MeV}$~\cite{baker1993}. } 
\label{fig:omp_12C}
\end{figure}

The solid curves in Fig.~\ref{fig:IDi_result} present 
the DWIA results with a shell-model wave function 
where the input parameters are same as those described in 
Sec~\ref{sec:DWIA_calculations}. 
The calculations underestimate all three spin-dependent $ID_i$ 
at $q \simeq 1.6~{\rm fm}^{-1}$, 
and the discrepancy in the momentum transfer dependence 
is evident in the spin-transverse $ID_p$ and $ID_n$. 
Thus, in Fig.~\ref{fig:IDi_par}, 
the sensitivity of the DWIA calculations for three spin-dependent $ID_i$ was investigated
for changes in the parameters. 
Note that the solid curves are the same as those in
Fig.~\ref{fig:IDi_result}. 
First, the OMP dependence of the calculations was examined 
by using other OMPs; three global OMPs (EDAD Fit~1--3) parameterized 
for $^{12}{\rm C}$--$^{208}{\rm Pb}$ in the proton energy range 
of $T_p=20$--$1040~{\rm MeV}$~\cite{hama1990,cooper1993} 
and the OMP obtained from proton elastic scattering data on 
$^{12}{\rm C}$ at $T_p=318~{\rm MeV}$ 
whose parameters are listed in Table~\ref{tab:omp_12C}~\cite{baker1993}. 
The radial dependences of these OMPs for the incident channel 
are shown in Fig.~\ref{fig:omp_12C}, and 
the DWIA results are shown in Fig.~\ref{fig:IDi_par}(a) by the bands. 
The OMP dependence of the spin-longitudinal $ID_q$ is small, 
whereas those of the spin-transverse $ID_p$ and $ID_n$ are significantly 
larger near the cross section minimum at $q \simeq 1.4~{\rm fm}^{-1}$. 
We also performed DWIA calculations using the neutron global 
OMPs for $^{12}{\rm C}$--$^{238}{\rm U}$ in the neutron energy range 
of $T_n=20$--$1000~{\rm MeV}$~\cite{qing1991} 
for the exit channel, and the results 
are plotted as the dashed curves in Fig.~\ref{fig:IDi_par}(a). 
The use of the neutron global OMPs gives 
larger values near the cross section minimum at 
$q \gtrsim 1.4~{\rm fm}^{-1}$ in the spin-transverse mode. 
However, neither the discrepancy in the angular distribution 
nor the underestimation in the cross section at large momentum transfers 
can be explained by the OMP uncertainties. 
\begin{table*}
\begin{center}
\begin{tabular}{lccc}\hline \hline
Potential & $V_i$ (MeV) & $r_i$ (fm) & $a_i$ (fm) \\ \hline
Real central ($i={\rm R}$) & -5.005 & 1.272 & 0.411 \\
Imaginary cetral ($i={\rm I}$) & -22.55 & 1.083 & 0.474 \\
Real spin-orbit ($i={\rm RSO}$) & -1.77 & 0.910 & 0.867 \\
Imaginary spin-orbit ($i={\rm ISO}$) & 2.71 & 0.909 & 0.467 \\
Coulomb ($i={\rm C}$) & & 1.24 & \\ \hline \hline
\end{tabular}
\end{center}
\caption{
The optical model parameters obtained from proton 
elastic scattering data on $^{12}{\rm C}$ at 
$T_p=318~{\rm MeV}$~\cite{baker1993}. 
The potential is defined by 
$U(r)=V_{\rm C}(r) + V_{\rm R} f_{\rm R}(r) + i V_{\rm I} f_{\rm I}(r) + 
[\hbar/(m_{\pi}c)]^2 (1/r) 
[V_{\rm RSO} (d/dr) f_{\rm RSO}(r) +
i V_{\rm ISO} (d/dr) f_{\rm ISO}(r)] 
(\mbox{\boldmath $\sigma$} \times \mbox{\boldmath $L$})$, 
where $V_{\rm C}$ is the coulomb potential for a uniformly 
charged sphere and $f_i(r)=[1+\exp\{(r-r_i A^{1/3})/a_i\}]^{-1}$. 
}
\label{tab:omp_12C}
\end{table*}

Second, the proton-particle and neutron-hole 
configuration dependences were investigated. 
The bands in Fig.~\ref{fig:IDi_par}(b) are DWIA results 
with other CKWFs based on the (8--16) 2BME and (8--16) POT 
interactions~\cite{cohen1965}. 
We also performed DWIA calculations for 
a pure $0p_{1/2}0p_{3/2}^{-1}$ transition from 
the Hartree-Fock (HF) state of $^{12}{\rm C}$ 
(the state fully occupying the $0s_{1/2}$ and $0p_{3/2}$ orbits), 
and the results are shown as the dashed curves. 
Table~\ref{tab:ckwf} summarizes the OBDMEs and $B$(GT) 
together with the corresponding normalization factors $N$ 
for the transition form factors. 
The configuration dependence is small for all three $ID_i$, and thus 
the discrepancy between the experimental and theoretical results 
is not resolved by considering the configuration dependence. 
\begin{table*}
\begin{center}
\begin{tabular}{ccccccc}\hline \hline
shell-model wave function & \multicolumn{4}{c}{OBDME} & $B$(GT) & $N$ \\ \hline
& $0p_{1/2}0p_{1/2}^{-1}$ & $0p_{1/2}0p_{3/2}^{-1}$ &
$0p_{3/2}0p_{1/2}^{-1}$ & $0p_{3/2}0p_{3/2}^{-1}$ & & \\ \hline
(6--16)2BME & 0.0859 & 0.6672 & 0.3228 & 0.0925 & 0.929 & 0.94 \\
(8--16)2BME & 0.0733 & 0.6915 & 0.3262 & 0.0822 & 0.992 & 0.88 \\
(8--16)POT  & 0.0582 & 0.6902 & 0.3393 & 0.0763 & 0.921 & 0.95 \\
Pure $0p_{1/2}0p_{3/2}^{-1}$ & 0.0 & 1.0 & 0.0 & 0.0 & 5.228 & 0.17
\\ \hline \hline
\end{tabular}
\end{center}
\caption{One-body density matrix elements and 
Gamow-Teller strengths for the 
$^{12}{\rm C}(p,n)^{12}{\rm N}({\rm g.s.},1^+)$ reaction
used in the DWIA calculations. 
The normalization factors for the transition form factors are also listed.}
\label{tab:ckwf}
\end{table*}

Finally, we investigated the dependence on the radial wave function. 
The dashed curves in Fig.~\ref{fig:IDi_par}(c) are DWIA results 
using a harmonic oscillator (HO) potential with a size parameter 
of $b=1.53~{\rm fm}^{-1}$. 
This parameter was obtained from an analysis of 
the electron scattering on $^{12}{\rm C}$ to the stretched 
$4^-$, $T=1$ state at $E_x=19.55~{\rm MeV}$~\cite{clausen1988} 
with the center-of-mass correction 
taken into account~\cite{comfort1981-2}. 
The results are almost the same as those for the WS potential. Thus 
the discrepancy between the experimental and theoretical results 
could not be explained by the radial wave function dependence. 

\begin{figure}[t]
\includegraphics[width=0.9\hsize,clip]{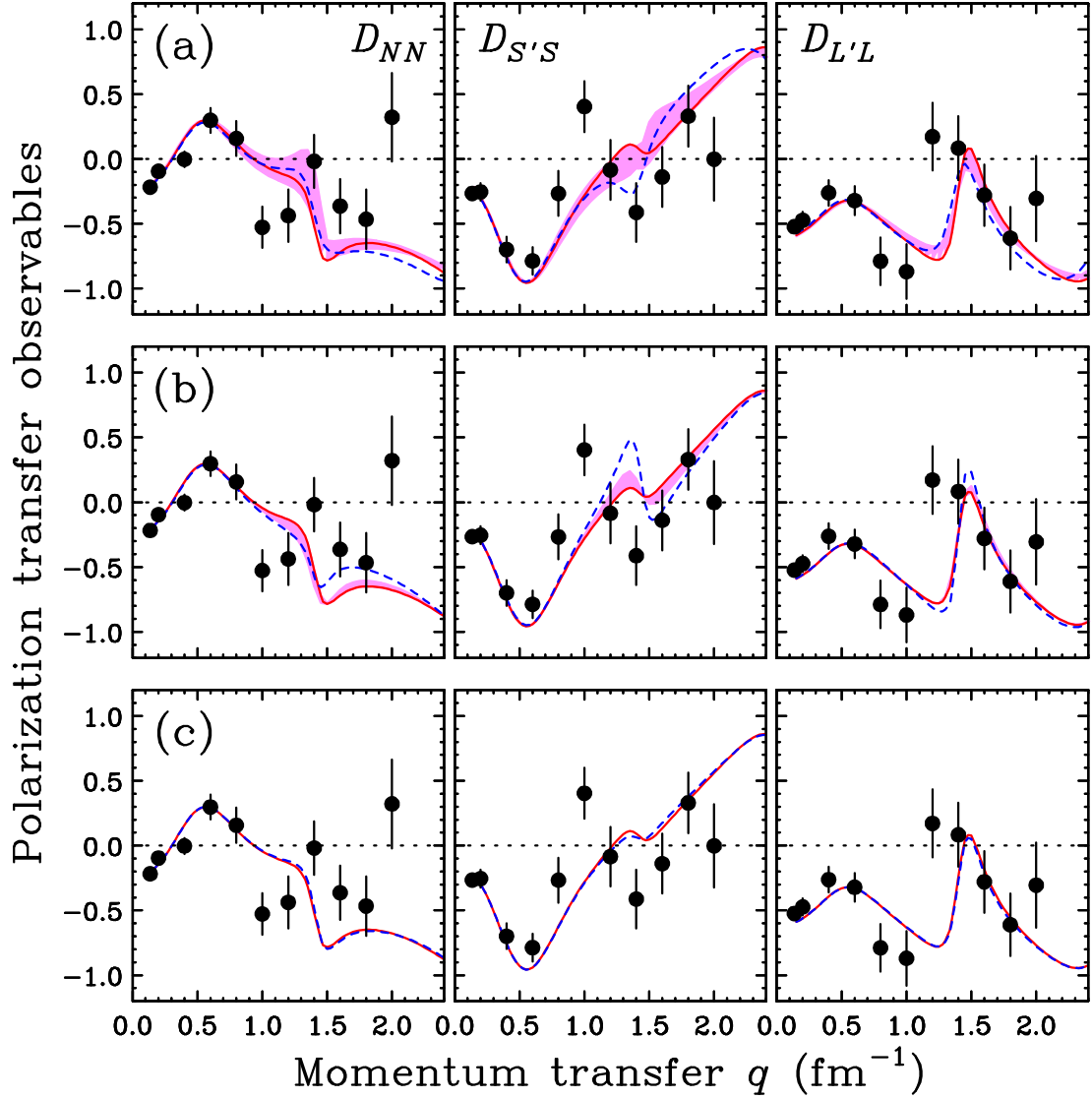}
\caption{(Color online) Same as in Fig.~\ref{fig:IDi_par} but for the 
diagonal polarization transfer observables: $D_{NN}$ (left panels), 
$D_{S'S}$ (middle panels), and $D_{L'L}$ (right panels).}
\label{fig:IDi_par1}
\end{figure}

Figure~\ref{fig:IDi_par1} represents the parameter dependence 
for the orthogonal components of the polarization transfer 
observables, $D_{NN}$ (left panels), $D_{S'S}$ (middle panels), 
and $D_{L'L}$ (right panels). 
Based on the calculations 
in Figs.~\ref{fig:IDi_par} and~\ref{fig:IDi_par1}, 
the experimental data 
at large momentum transfers cannot be reproduced 
within the framework of the DWIA employing shell-model wave functions. 
Therefore, in the following section, 
nuclear correlation effects beyond the shell model are investigated. 

\section{\label{sec:discussion}DISCUSSION}
In this section, the experimental 
spin-dependent polarized cross sections are compared
with the DWIA calculations using RPA response functions in order to 
investigate the nuclear correlation effects beyond the shell model. 

\subsection{\label{sec:DWIA+RPA_calculations}
DWIA+RPA calculations}
We performed DWIA+RPA calculations using the computer code 
{\sc crdw}~\cite{kawahigashi2001}. 
The formalism of the calculations is discussed in 
Refs.~\cite{nishida1995,kawahigashi2001}. 
The spin--isospin response functions were calculated in a
continuum RPA including the $\Delta$ degrees of freedom. 
We further utilized a ring approximation~\cite{kawahigashi2001}, 
and used the $\pi + \rho + g'$ model interaction 
for the effective interaction, 
which is expressed as~\cite{ichimura2006}
\begin{equation}
V_{\rm eff}(\mbox{\boldmath $q$},\omega)= 
V_L(\mbox{\boldmath $q$},\omega) +
V_T(\mbox{\boldmath $q$},\omega),
\end{equation}
where $V_L$ and $V_T$ are the spin-longitudinal and spin-transverse 
effective interactions, respectively. 
They are determined by 
the pion and rho-meson exchange interactions 
and the Landau-Migdal (LM) interaction specified by 
the LM parameters, $g'_{NN}$, $g'_{N\Delta}$, and $g'_{\Delta\Delta}$, as 
\begin{equation}
\begin{array}{ccl}
V_L(\mbox{\boldmath $q$},\omega) & = &
{\displaystyle \frac{f^2_{\pi NN}}{m^2_{\pi}}}
\left(g'_{NN}+{\displaystyle\frac{q^2}{\omega^2-q^2-m^2_{\pi}}}\Gamma^2_{\pi NN}
(q,\omega)\right)
(\mbox{\boldmath $\sigma$}_1 \cdot \mbox{\boldmath $\hat{q}$})
(\mbox{\boldmath $\sigma$}_2 \cdot \mbox{\boldmath $\hat{q}$})
(\mbox{\boldmath $\tau$}_1 \cdot \mbox{\boldmath $\tau$}_2) \\[15pt]
& & +
{\displaystyle \frac{f_{\pi NN}f_{\pi N\Delta}}{m^2_{\pi}}}
\left(g'_{N\Delta}+{\displaystyle\frac{q^2}{\omega^2-q^2-m^2_{\pi}}}
\Gamma_{\pi NN}(q,\omega)
\Gamma_{\pi N\Delta}
(q,\omega)\right)\\[15pt]
& & \times
\left[\left\{
(\mbox{\boldmath $\sigma$}_1 \cdot \mbox{\boldmath $\hat{q}$})
(\mbox{\boldmath $S$}_2 \cdot \mbox{\boldmath $\hat{q}$})
(\mbox{\boldmath $\tau$}_1 \cdot \mbox{\boldmath $T$}_2)
+(1 \leftrightarrow 2)
\right\}+ {\rm h.c.}\right]\\[10pt]
& & +
{\displaystyle \frac{f^2_{\pi N\Delta}}{m^2_{\pi}}}
\left(g'_{\Delta\Delta}+{\displaystyle\frac{q^2}{\omega^2-q^2-m^2_{\pi}}}
\Gamma^2_{\pi N\Delta}
(q,\omega)\right)\\[15pt]
& & \times 
\left[\left\{
(\mbox{\boldmath $S$}_1 \cdot \mbox{\boldmath $\hat{q}$})
(\mbox{\boldmath $S$}_2^\dagger \cdot \mbox{\boldmath $\hat{q}$})
(\mbox{\boldmath $T$}_1 \cdot \mbox{\boldmath $T$}_2^\dagger)+
(\mbox{\boldmath $S$}_1 \cdot \mbox{\boldmath $\hat{q}$})
(\mbox{\boldmath $S$}_2 \cdot \mbox{\boldmath $\hat{q}$})
(\mbox{\boldmath $T$}_1 \cdot \mbox{\boldmath $T$}_2)
\right\} + {\rm h.c.}\right], 
\end{array}
\end{equation}
and 
\begin{equation}
\begin{array}{ccl}
V_T(\mbox{\boldmath $q$},\omega) & = &
{\displaystyle \frac{f^2_{\pi NN}}{m^2_{\pi}}}
\left(g'_{NN}+C_{\rho}{\displaystyle\frac{q^2}{\omega^2-q^2-m^2_{\rho}}}\Gamma^2_{\rho NN}
(q,\omega)\right)\\[15pt]
& & \times
(\mbox{\boldmath $\sigma$}_1 \times \mbox{\boldmath $\hat{q}$})
(\mbox{\boldmath $\sigma$}_2 \times \mbox{\boldmath $\hat{q}$})
(\mbox{\boldmath $\tau$}_1 \cdot \mbox{\boldmath $\tau$}_2) \\[10pt]
& & +
{\displaystyle \frac{f_{\pi NN}f_{\pi N\Delta}}{m^2_{\pi}}}
\left(g'_{N\Delta}+C_{\rho}{\displaystyle\frac{q^2}{\omega^2-q^2-m^2_{\rho}}}
\Gamma_{\rho NN}(q,\omega)
\Gamma_{\rho N\Delta}
(q,\omega)\right)\\[15pt]
& & \times
\left[\left\{
(\mbox{\boldmath $\sigma$}_1 \times \mbox{\boldmath $\hat{q}$})
(\mbox{\boldmath $S$}_2 \times \mbox{\boldmath $\hat{q}$})
(\mbox{\boldmath $\tau$}_1 \cdot \mbox{\boldmath $T$}_2)
+(1 \leftrightarrow 2)
\right\} + {\rm h.c.} \right]\\[10pt]
& & +
{\displaystyle \frac{f^2_{\pi N\Delta}}{m^2_{\pi}}}
\left(g'_{\Delta\Delta}+C_{\rho}{\displaystyle\frac{q^2}{\omega^2-q^2-m^2_{\rho}}}
\Gamma^2_{\rho N\Delta}
(q,\omega)\right)\\[15pt]
& & \times
\left[
\left\{
(\mbox{\boldmath $S$}_1 \times \mbox{\boldmath $\hat{q}$})
(\mbox{\boldmath $S$}_2^\dagger \times \mbox{\boldmath $\hat{q}$})
(\mbox{\boldmath $T$}_1 \cdot \mbox{\boldmath $T$}_2^\dagger) + 
(\mbox{\boldmath $S$}_1 \times \mbox{\boldmath $\hat{q}$})
(\mbox{\boldmath $S$}_2 \times \mbox{\boldmath $\hat{q}$})
(\mbox{\boldmath $T$}_1 \cdot \mbox{\boldmath $T$}_2)
\right\} + {\rm h.c.}
\right],
\end{array}
\end{equation}
where $m_{\pi}$ and $m_{\rho}$ are the pion and rho-meson masses, 
$\sigma$ and $\tau$ are the spin and isospin operators of the nucleon $N$, 
and $\mbox{\boldmath $S$}$ and $\mbox{\boldmath $T$}$ are the 
spin and isospin transition operators from $N$ to $\Delta$. 
The coupling constants and meson parameters for the 
pion and rho-meson exchange interactions from a Bonn potential were used, 
which treats $\Delta$ explicitly~\cite{machleidt1987}. 
The LM interaction effectively represents the short range correlations 
and the exchange terms in the RPA, and 
the LM parameters have been estimated to be 
$g'_{NN}=0.65 \pm 0.15$ and $g'_{N\Delta}=0.35 \pm 0.15$~\cite{wakasa2005-2} 
by using the peak position of the GT giant resonance and the GT quenching 
factor at $q=0~{\rm fm}^{-1}$~\cite{wakasa1997,yako2005}, 
as well as the isovector spin-longitudinal polarized 
cross section in the QES process at 
$q \simeq 1.7~{\rm fm}^{-1}$~\cite{wakasa2004-2}. 
We fixed $g'_{\Delta \Delta}=0.5$~\cite{dickhoff1981} 
because the $g'_{\Delta \Delta}$ 
dependence in the results is weak. 
The response functions are normalized to reproduce the 
experimental $B({\rm GT})$. 

The ground state of $^{12}{\rm C}$ was assumed to be a HF state. 
However, as seen in Table~\ref{tab:ckwf}, 
the ground state correlation which is included in 
the shell-model calculations plays an important role 
in reproducing the experimental $B({\rm GT})$ value of 0.873. 
In order to include the shell-model (configuration-mixing) 
effects effectively, 
we used much smaller normalization factors $N$ than those of 
the shell-model calculations, namely $N=0.28$ and $N=0.17$ 
in the calculations with and without RPA correlations, respectively. 

The nonlocality of the nuclear mean field was treated using the 
local effective mass approximation~\cite{ichimura2006} in the form
\begin{equation}
m^*(r)=m_N-\frac{f_{\rm WS}(r)}{f_{\rm WS}(0)}(m_N-m^*(0)), 
\end{equation}
where $m_N$ is the nucleon mass and $f_{\rm WS}(r)$ is a WS radial form. 
Here we adapted the standard value of 
$m^*(0)=0.7m_N$~\cite{giai1983,mahaux1988}. 

We used the same OMPs and single-particle wave functions 
as described in Sec.~\ref{sec:DWIA_calculations}. 
The $NN$ $t$-matrix parameterized by Franey and Love at 325~MeV 
was used, and the exchange terms were approximated by 
contact terms following the prescription 
by Love and Franey~\cite{love1981}. 

\begin{figure}[t]
\includegraphics[width=0.9\hsize,clip]{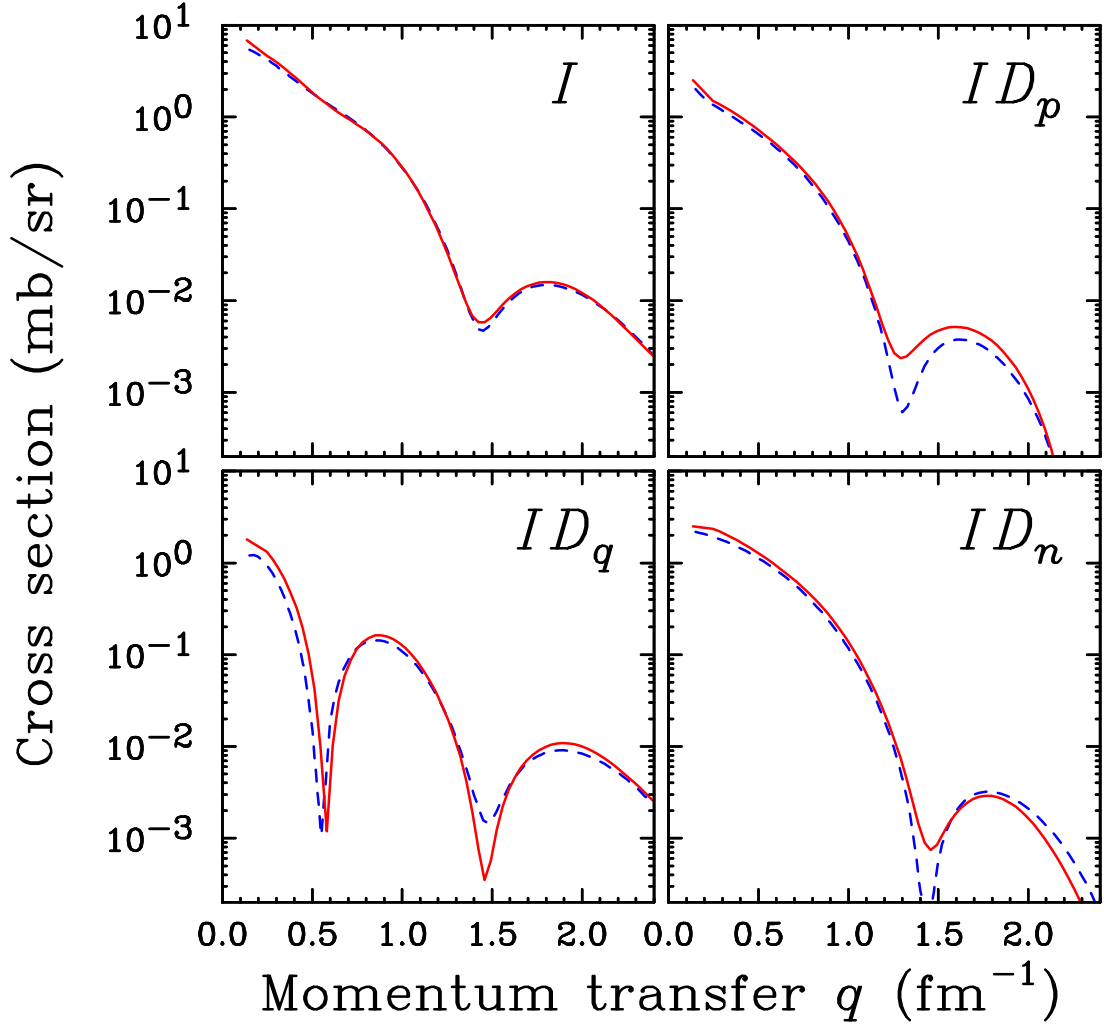}
\caption{(Color online) Comparison between calculations 
using the computer codes {\sc crdw} and {\sc dw81} 
for the cross section $I$ and 
polarized cross sections, $ID_q$, $ID_p$, and $ID_n$. 
The solid and dashed curves represent 
DWIA calculations using {\sc crdw} 
and {\sc dw81}, respectively. 
The normalization factor $N$ of 0.17 is common for both calculations.} 
\label{fig:crdw_vs_dw81}
\end{figure}

In Fig.~\ref{fig:crdw_vs_dw81}, the consistency 
between theoretical results using the computer codes 
{\sc crdw} and {\sc dw81} is checked. 
The solid curves are the {\sc crdw} results 
with a free response function employing $m^*(0)=m_N$, 
whereas the dashed curves are 
the {\sc dw81} results with the corresponding wave function 
as for the dashed curves in Fig.~\ref{fig:IDi_par}(b). 
The normalization factor $N$ of 0.17 is common, and 
both calculations for the unpolarized cross section 
$I$ are in good agreement with each other. 
The theoretical results for $ID_i$ 
show good consistency except near 
the cross section minimum at $q \simeq 1.4~{\rm fm}^{-1}$. 
Therefore, in the next subsection, 
the experimental data are compared
with the DWIA calculations using 
{\sc crdw} in order to investigate 
nuclear structure effects 
which could not be included in the preceding calculations, 
such as RPA correlations. 

\subsection{Comparison with DWIA+RPA calculations\label{sec:comparison}}
\begin{figure}[t]
\includegraphics[width=0.9\hsize,clip]{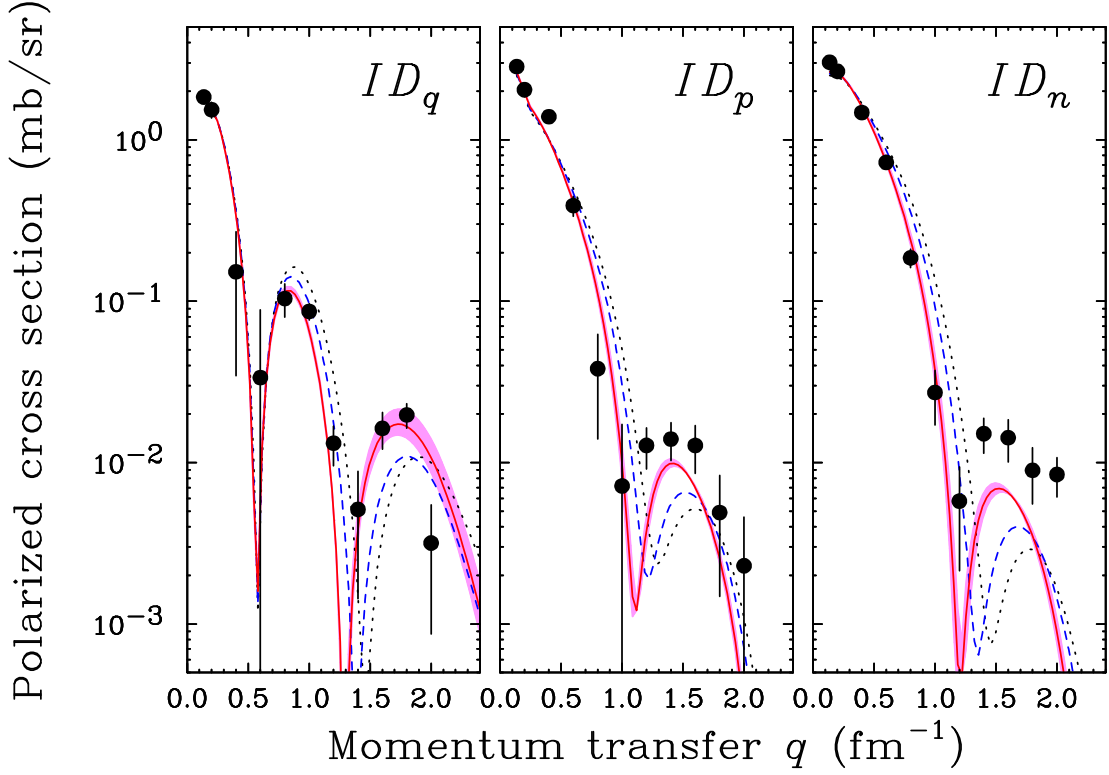}
\caption{(Color online) 
Comparison between experimental and theoretical results of 
polarized cross sections $ID_q$, $ID_p$, and $ID_n$ for 
$^{12}{\rm C}(p,n)^{12}{\rm N}({\rm g.s.},1^+)$ at 
$T_p=296~{\rm MeV}$. 
The dotted and dashed curves present the DWIA results 
with a free response function using $m^*(0)=m_N$ and 
$m^*(0)=0.7m_N$, respectively, and $N=0.17$. 
The solid curves denote DWIA results for an RPA 
response function with 
$(g'_{NN},g'_{N\Delta},g'_{\Delta \Delta})=(0.65, 0.35, 0.50)$ 
, $m^*(0)=0.7m_N$ and $N=0.28$. 
The bands are the $g'_{NN}$ and $g'_{N\Delta}$ dependences of 
the DWIA results with $g'_{NN}=0.65 \pm 0.15$ and 
$g'_{N\Delta} = 0.35 \pm 0.15$. } 
\label{fig:crdw_result1}
\end{figure}

First, the nonlocality of the nuclear mean field was investigated. 
The dotted and dashed curves in Fig.~\ref{fig:crdw_result1} 
show the DWIA results with a free response function 
using $m^*(0)=m_N$ and $m^*(0)=0.7 m_N$, respectively, 
and $N=0.17$. 
The angular distributions of all three spin-dependent $ID_i$ curves shift to 
lower momentum transfer due to the nonlocality of the 
nuclear mean field, so that the agreement with the data is improved. 
This shift arises because the transition form factor moves 
outward due to the Perey effect~\cite{perey1962}. 
However, there remains a discrepancy between the experimental and 
theoretical results at around $q \simeq 1.6~{\rm fm}^{-1}$. 

Next we considered the nuclear correlation effects in the RPA. 
The solid curves in Fig.~\ref{fig:crdw_result1} 
show the results of DWIA+RPA calculations 
using $g'_{NN}=0.65$, $g'_{N\Delta}=0.35$, and $m^*(0)=0.7 m_N$ 
with $N=0.28$. 
The bands represent the $g'_{NN}$ and $g'_{N\Delta}$ dependences 
with $g'_{NN}=0.65 \pm 0.15$ and $g'_{N\Delta}=0.35 \pm 0.15$. 
In the continuum RPA, the GT state couples to particle-unbound $1^+$ 
states, which shifts the response function in coordinate space 
to larger $r$ values. 
Thus the angular distributions further shift to lower momentum
transfer. 
Furthermore, the RPA correlation enhances all three $ID_i$ 
at large momentum transfers of $q \simeq 1.6~{\rm fm}^{-1}$, 
improving the agreement with the data. 
In the analysis of QES, the spin-transverse $ID_p$ and $ID_n$ are 
quenched due to the repulsion of the spin-transverse interaction 
$V_T$~\cite{wakasa1999,wakasa2004-2}. 
However, in the analysis shown in Fig.~\ref{fig:crdw_result1}, 
the transition form factors are normalized to reproduce $B$(GT). 
This means that the quenching due to the repulsive LM interaction is 
effectively included through the normalization factor 
$N$~\cite{wakasa2007-2}. 
Therefore, in Fig.~\ref{fig:crdw_result1}, 
the attractive rho-meson exchange effects are 
seen as an enhancement of $ID_p$ and $ID_n$  
at $q \simeq 1.6~{\rm fm}^{-1}$ 
Also note that the modification of the momentum transfer dependences 
(due to the shape change of the response functions in $r$-space 
which is not included through the normalization) 
is important for the magnitude of $ID_i$. 

\begin{figure}[t]
\includegraphics[width=0.9\hsize,clip]{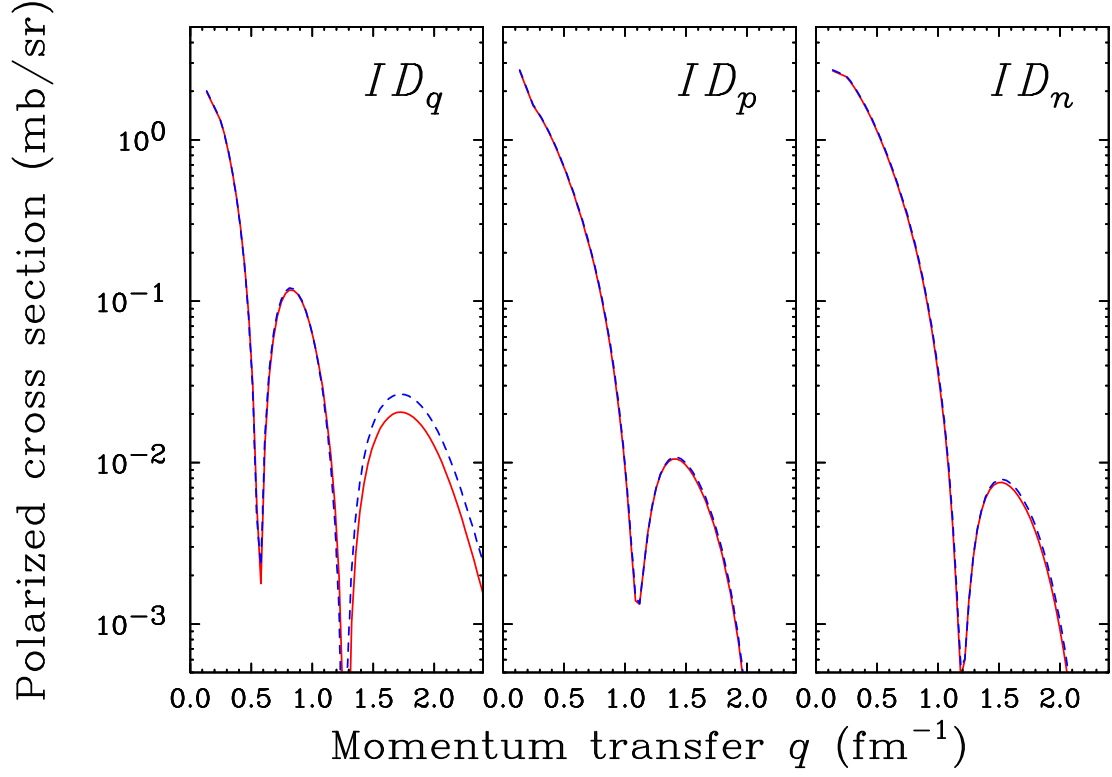}
\caption{(Color online) 
DWIA predictions for $^{12}{\rm C}(p,n)^{12}{\rm N}({\rm g.s.},1^+)$ 
at $T_p=296~{\rm MeV}$. 
The dashed and solid curves are the DWIA results 
with the RPA response functions using the LM 
parameters with and without the dipole form factors, respectively. 
} 
\label{fig:formfactor}
\end{figure}

In some theoretical studies~\cite{oset1982,alberico1982,speth1980}, 
the LM parameters are taken to have 
the momentum- and energy-transfer dependence. 
We investigated this effect on $ID_i$ by 
using the effective interaction $V_{\rm eff}$ 
by Alberico {\it et al.}~\cite{alberico1982}, 
in which the LM parameters have the momentum- and 
energy-transfer dependence with the dipole form factors. 
The DWIA+RPA results with this effective interaction 
are shown in Fig.~\ref{fig:formfactor}. 
The dashed and solid curves correspond to the calculations 
with and without the dipole form factors, respectively. 
We note that the coupling constants and meson parameters 
are slightly different from those in Ref.~\cite{machleidt1987}, 
and thus the results without the dipole form factors are also 
slightly different from those shown in Fig.~\ref{fig:crdw_result1}. 
For the spin-longitudinal mode, 
the use of the dipole form factors enhances $ID_q$ 
at large momentum transfers of $q \simeq 1.6~{\rm fm}^{-1}$, 
which is due to the more attractive spin-longitudinal 
interaction $V_L$. 
However, the effect is small, and 
the same results can be achieved by using 
smaller and reasonable LM parameters of 
$g'_{NN} \simeq 0.55$ and $g'_{N\Delta} \simeq 0.30$. 
For the spin-transverse mode, 
the form factor effects are very small 
in both $ID_p$ and $ID_n$ 
because of the insensitivity to the spin-transverse 
interaction $V_T$~\cite{ichimura2006}. 

\begin{figure}[t]
\includegraphics[width=0.9\hsize,clip]{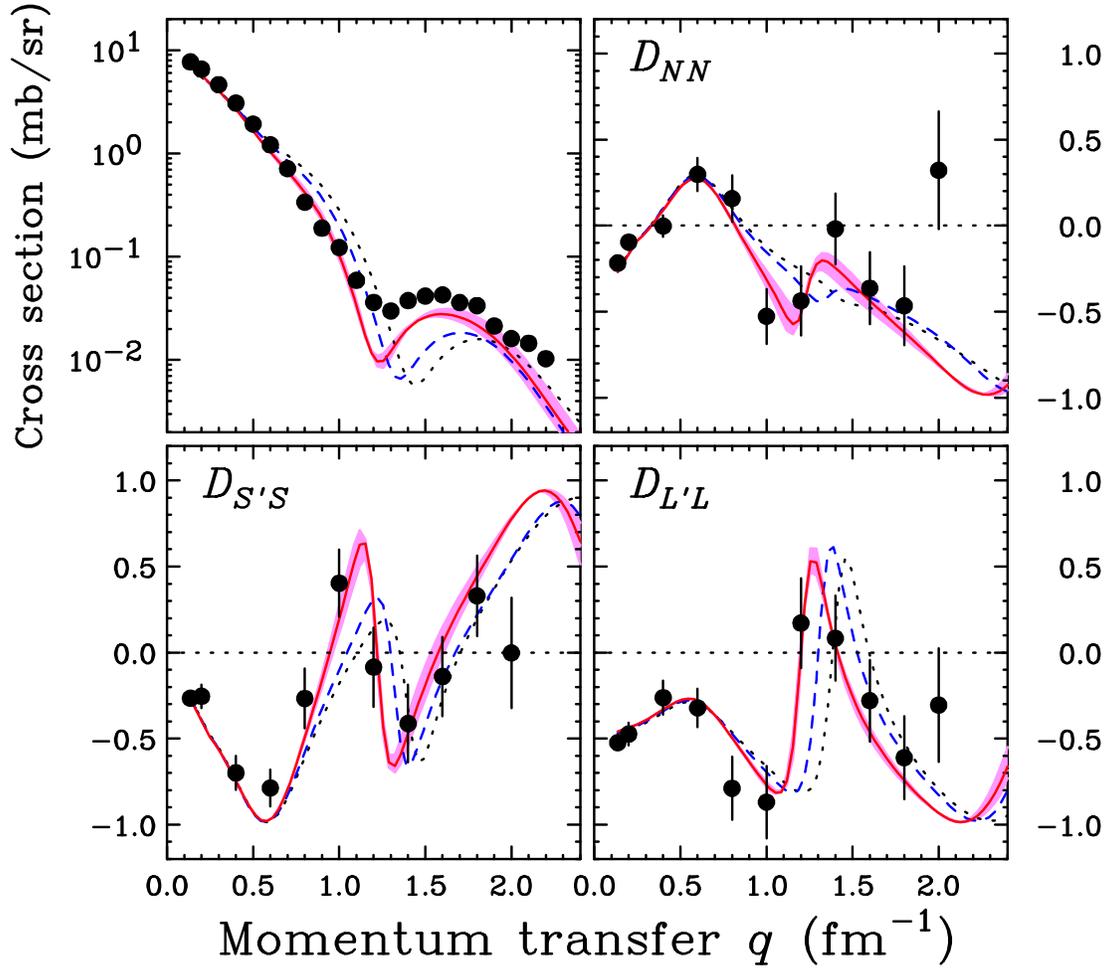}
\caption{(Color online) 
Same as in Fig.~\ref{fig:crdw_result1} but for the cross section 
and diagonal polarization transfer observables, 
$D_{NN}$, $D_{S'S}$, and $D_{L'L}$. }
\label{fig:crdw_result2}
\end{figure}

Figure~\ref{fig:crdw_result2} compares the 
experimental and theoretical results for the cross section and 
orthogonal components of the polarization transfer observables, 
$D_{NN}$, $D_{S'S}$, and $D_{L'L}$. 
These quantities are better reproduced 
by considering RPA correlations together with the nonlocality 
of the nuclear mean field, 
particularly for the momentum transfer dependences. 
From the analyses in Figs.~\ref{fig:crdw_result1} 
and \ref{fig:crdw_result2}, 
we conclude that our data support 
the existence of pionic and rho-mesonic correlations in nuclei 
at large momentum transfers. 

\section{\label{sec:final_remark} FINAL REMARKS}

The DWIA calculations including RPA correlations 
reproduce the experimental data for the spin-longitudinal cross section 
$ID_q$ and 
give improved descriptions of the spin-transverse cross sections, 
$ID_p$ and $ID_n$. 
However, the experimental values of $ID_p$ and $ID_n$ remain larger 
than the calculated values at momentum transfers of 
$q \simeq 1.6~{\rm fm}^{-1}$ 
by factors of about 1.4 and 2.0, respectively. 
The magnitudes of the observed enhancements are 
significantly different for $ID_p$ and $ID_n$. 
Since the spin-transverse response is common for $ID_p$ and $ID_n$, 
medium modifications of the effective $NN$ interaction 
are considered as a possible reason for the observed enhancements. 
Such modifications have been discussed using 
data for the stretched state excitations 
in Refs.~\cite{wakasa2007-1,stephenson1997,yako1998}, 
where it is reported that 
the experimental values of $ID_n$ at large momentum transfers are 
increased in magnitude by a factor of about 1.5, indicating an enhancement in
the corresponding $NN$ scattering amplitude $B$. 
Thus, a larger value of $B$ might be responsible for the enhancement 
in the experimental values of $ID_n$ from Eq.~(\ref{eq:fia}). 
However, since no modification of the $NN$ amplitude $F$ is 
observed in the stretched state excitations, 
the enhancement in $ID_p$ cannot 
be explained by modifications of the $NN$ interaction. 
We note that medium modifications of the effective $NN$ interaction 
have been also discussed in the relativistic framework~\cite{hillhouse1998}. 
If one takes into account 
the increase in $B$, the enhancement of $ID_n$ 
is reduced to 1.3, which is 
almost same as that of $ID_p$. 
Consequently the experimental values of $ID_p$ and $ID_n$ may be 
enhanced by a common mechanism, such as higher order (e.g., 2p-2h) 
configuration mixing~\cite{alberico1984,adams1994}. 
Thus more comprehensive and detailed theoretical analyses are needed 
including effects of higher order configuration mixing 
and medium modifications of the effective $NN$ interaction. 

\section{\label{sec:summary_and_conclusion}
SUMMARY AND CONCLUSION}
We have measured differential cross sections and 
complete sets of polarization observables 
for the $^{12}{\rm C}(\vec{p},\vec{n})^{12}{\rm N}({\rm g.s.,}1^+)$ 
reaction at $T_p=296~{\rm MeV}$ with momentum transfers of $q=0.1$ to $2.2~{\rm fm}^{-1}$ 
in order to investigate nuclear correlation effects inside the nuclei. 
The experimental polarized cross sections $ID_i$ have been compared 
with DWIA calculations employing shell-model wave functions. 
For all three spin-dependent $ID_i$, 
a significant difference in the momentum-transfer dependence 
and an enhancement around $q \simeq 1.6~{\rm fm}^{-1}$ were 
observed when compared to calculations. 
The use of a local effective mass of $m^*(0) = 0.7 m_N$ improves 
the agreement with the data, but still underestimates 
the cross section at around $q \simeq 1.6~{\rm fm}^{-1}$. 
These underestimations for all three spin-dependent $ID_i$ 
are partly resolved by DWIA calculations 
employing an RPA response function with $g'_{NN}=0.65$, 
$g'_{N\Delta}=0.35$, and $m^* (0) =0.7 m_N$, supporting 
the existence of pionic and rho-mesonic correlations in the nuclei. 
This finding is the first indication for observing 
pionic and rho-mesonic correlation effects separately. 
To understand the nuclear correlation effects quantitatively, 
theoretical analyses are required that include 
effects of higher order configuration mixing and 
medium modifications in the effective $NN$ interaction. 

\begin{acknowledgments}
We thank the RCNP cyclotron crew for providing a good quality 
beam for this experiment, 
performed under Program Number E256. 
This work was supported by Grants-in-Aid for 
Scientific Research Nos. 14702005 and 16654064 from the Ministry of 
Education, Culture, Sports, Science, and Technology of Japan. 
\end{acknowledgments}


\newpage 

\newpage

\end{document}